\documentclass[10pt,twoside,a4paper]{article}

\usepackage{amsmath}
\usepackage{amssymb}
\usepackage{amsthm}
\usepackage{amsfonts}
\usepackage{amstext}
\usepackage{amsopn}
\usepackage{amsxtra}
\usepackage{graphicx}
\usepackage[latin1]{inputenc}
\newcommand\R{{\ensuremath {\mathbb R} }}
\newcommand\C{{\ensuremath {\mathbb C} }}

\newcommand\Z{{\ensuremath {\mathbb Z} }}

\newcommand\1{{\ensuremath {1}}} %\mathds

\renewcommand\phi{\varphi}

\newcommand{\alp}{\boldsymbol{\alpha}}

\newcommand{\gH}{\mathfrak{H}}
\newcommand{\gS}{\mathfrak{S}}

\newcommand{\wto}{\rightharpoonup}
\newcommand{\cP}{\mathcal{P}}
\newcommand{\cD}{\mathcal{D}}

\newcommand{\cB}{\mathcal{B}}

\newcommand\ii{{\ensuremath {\infty}}}
\newcommand\pscal[1]{{\ensuremath{\left\langle #1 \right\rangle}}}
\newcommand{\norm}[1]{ \left| \! \left| #1 \right| \! \right| }

\def\tr{\mathop{\rm tr}\nolimits} % trace
\def\str{\mathop{\rm tr}_{\cP^0_-}\nolimits} % supertrace
\newcommand{\E}{\mathcal{E}}

\newcommand{\cC}{\mathcal{C}}
\newcommand{\cQ}{\mathcal{Q}}
\newcommand{\cK}{\mathcal{K}}

\newtheorem{theorem}{Theorem}
\newtheorem{lemma}{Lemma}
\newtheorem{corollary}[lemma]{Corollary}
\newtheorem{proposition}[lemma]{Proposition}

\newtheorem{remark}{Remark}

\title{\bf Existence of Atoms and Molecules in the Mean-Field Approximation of No-Photon Quantum Electrodynamics}
\author{\bf Christian HAINZL$^1$, Mathieu LEWIN$^2$ \& \'Eric S\'ER\'E$^3$}
\date{\today}
\begin{document}
\begin{center}
  \Large \textbf{Existence of Atoms and Molecules in the Mean-Field Approximation of No-Photon Quantum Electrodynamics\footnote{February 19, 2008. Final version to appear in \textit{Arch. Rat. Mech. Anal.} \copyright\,2008 by the authors. This work may be reproduced, in its entirety, for non-commercial purposes.}}
\end{center}

\smallskip

\begin{center}
 \large Christian HAINZL$^a$, Mathieu LEWIN$^b$ \& \'Eric S\'ER\'E$^c$
\end{center}

\begin{center}
\footnotesize $^a$ Department of Mathematics, University of Alabama at Birmingham. Campbell Hall. 1300 University Boulevard. Birmingham, Al-35294 USA.\\ E-mail: \texttt{hainzl@math.uab.edu}

\medskip

$^b$ CNRS \& Department of Mathematics (CNRS UMR8088), Université de Cergy-Pontoise. 2, avenue Adolphe Chauvin. 95302 Cergy-Pontoise Cedex FRANCE.\\ Email: \texttt{Mathieu.Lewin@math.cnrs.fr}

\medskip

$^c$ CEREMADE (CNRS UMR 7534), Université Paris-Dauphine. Place du Maréchal De Lattre De Tassigny. 75775 Paris Cedex 16 FRANCE.\\ E-mail: \texttt{sere@ceremade.dauphine.fr}
\end{center}

\smallskip

\begin{abstract}
The Bogoliubov-Dirac-Fock (BDF) model is the mean-field approximation of no-photon Quantum Electrodynamics. 
The present paper is devoted to the study of the minimization of the BDF energy functional \emph{under a charge constraint}. An associated minimizer, if it exists, will usually represent the ground state of a system of $N$ electrons interacting with the Dirac sea, in an external electrostatic field generated by one or several fixed nuclei. We prove that such a minimizer exists when a binding (HVZ-type) condition holds. We also derive, study and interpret the equation satisfied by such a minimizer.

Finally, we provide two regimes in which the binding condition is fulfilled, obtaining the existence of a minimizer in these cases. The first is the weak coupling regime for which the coupling constant $\alpha$ is small whereas $\alpha Z$ and the particle number $N$ are fixed. The second is the non-relativistic regime in which the speed of light tends to infinity (or equivalently $\alpha$ tends to zero) and $Z$, $N$ are fixed. We also prove that the electronic solution converges in the non-relativistic limit towards a Hartree-Fock ground state.
\end{abstract}

\section{Introduction}
The relativistic quantum theory of electrons and positrons is based 
on the free Dirac operator \cite{Dirac-28}, which is defined by
\begin{equation}
D^0=-i\hbar c\sum_{k=1}^3\alpha_k\partial_k+mc^2\beta :=-i\hbar c\,\alp\cdot \nabla+m c^2\beta
\label{dirac_free}
\end{equation}
$$\text{where}\quad \alp=(\alpha_1,\alpha_2,\alpha_3)\quad \text{and}\quad
\beta=\left(\begin{matrix}
I_2 & 0\\ 0 & -I_2\\
\end{matrix}\right),\qquad
\alpha_k=\left(\begin{matrix}
0 & \sigma_k\\ \sigma_k & 0\\
\end{matrix}\right),$$
$$\sigma_1=\left(\begin{matrix}
0 & 1\\ 1 & 0\\
\end{matrix}\right), \qquad
\sigma_2=\left(\begin{matrix}
  0 & -i\\ i & 0\\
\end{matrix}\right), \qquad
\sigma_3=\left(\begin{matrix}
1 & 0\\ 0 & -1\\
\end{matrix}\right).$$
We follow here mainly the notation of Thaller's book \cite{Thaller}. In \eqref{dirac_free}, $\hbar$ is Planck's constant, $c$ is the speed of light and $m$ is the mass of a free electron. For the sake of simplicity, we shall use in the following a system of units such that $\hbar=m=1$. Unless otherwise specified, we shall also assume that $c=1$, in which case an additional parameter will appear in front of the interaction potentials, $\alpha=e^2$, where $-e$ is the bare charge of a free electron.

The operator $D^0$ acts on $4$-spinors, i.e. functions $\psi\in 
L^2(\R^3,\C^4)$. It is self-adjoint on $L^2(\R^3,\C^4)$, 
with domain $H^1(\R^3,\C^4)$ and form-domain $H^{1/2}(\R^3,\C^4)$. 
Moreover, it is defined to ensure
$(D^0)^2=-\Delta+1$.
The spectrum of $D^0$ is $\sigma(D^0)=(-\ii,-1]\cup [1,\ii)$. 
In what follows, the projector associated with the negative (resp. positive) part of 
the spectrum of $D^0$ will be denoted by $P^0_-$ (resp. $P^0_+$):
$$P^0_-:=\chi_{(-\ii,0)}(D^0),\qquad P^0_+:=\chi_{(0,+\ii)}(D^0).$$
We then have $D^0P^0_-=P^0_-D^0=-|D^0|\,P^0_-$ and $D^0P^0_+=P^0_+D^0=|D^0|\,P^0_+$.

Compared with the non-relativistic (Schrödinger) models in which $-\Delta/2$ appears instead of $D^0$, the main unusual feature of the relativistic theories is that $\sigma(D^0)$ is not bounded from below. Indeed the free Dirac operator \eqref{dirac_free} was proposed by Dirac in 1928 \cite{Dirac-28} to describe the energy of a free relativistic spin-$1/2$ particle like an electron. In order to explain why negative energy electrons are never observed, Dirac  made the assumption \cite{Dirac-30,Dirac-34a,Dirac-34b} that the vacuum is filled with infinitely many virtual electrons occupying all the negative energy states so that, due to the Pauli principle, a physical free electron cannot have a negative energy. This model is commonly called the \emph{Dirac sea}. Mathematically,  the free vacuum is identified with the projector $P^0_-$.

With this interpretation, Dirac was able to conjecture the existence of the positron (the anti-electron,  which has a positive charge), which is seen as a hole in the vacuum and was discovered in 1932 by Anderson \cite{Anderson-33}. He also predicted interesting new physical features as a consequence of his theory \cite{Dirac-30,Dirac-34a,Dirac-34b}, which were experimentally confirmed later. First, the virtual electrons of the Dirac sea can feel an external field and they will react to this field accordingly, i.e. the vacuum will become \emph{polarized}.  From the experimental viewpoint, vacuum polarization plays a rather small role for the calculation of the Lamb shift of hydrogen but it is important for high-$Z$ atoms \cite{Mohr-Plunien-Soff-98} and it is even a crucial physical effect for muonic atoms \cite{Foldy-Eriksen-54,Glauber-60}. Second, in the presence of strong external fields, the vacuum can acquire a nonzero charge, a phenomenon which is related to the spontaneous creation of electron-positron pairs \cite{Nenciu,RG,RGA,RMG}. 

On the other hand, many models which are commonly used to describe relativistic particles do \emph{not} take the vacuum polarization effects into account. This is for instance the case of the (mean-field) Dirac-Fock theory which is the relativistic counterpart of the well-known Hartree-Fock model and was proposed by Swirles \cite{Swirles-35}. The Dirac-Fock model suffers from an important defect: the corresponding energy is not bounded from below, contrary to the Hartree-Fock case, and this leads to
important computational difficulties (see \cite{Chaix} for a discussion and detailed
references). From the mathematical viewpoint, one can 
prove that the Dirac-Fock functional has critical points which are solutions of the Dirac-Fock equations \cite{ES1,P}, but these critical points have an infinite Morse index, and the rigorous definition of a ground 
state is delicate \cite{ES2,ES3}. 

It was proposed by Chaix and Iracane \cite{CI} that these difficulties could be overcome by incorporating the vacuum polarization effects in the theory, i.e. by considering the coupled system `Dirac sea + real electrons' instead of the electrons alone. Starting from Quantum Electrodynamics (QED) and neglecting photons, they derived a model called Bogoliubov-Dirac-Fock (BDF), in which the real particles are coupled to the Dirac sea. The main advantage of this theory is that the energy of the model is now bounded below, leading to a clear definition of the ground state. 

The Chaix-Iracane model was first mathematically studied in the free case by Chaix, Iracane and Lions in \cite{CIL} and then by Bach, Barbaroux, Helffer and Siedentop in \cite{BBHS}. The external field case was rigorously defined and studied by the authors of the present paper in \cite{HLS1,HLS2}. Chaix and Iracane derived their functional under Dirac's assumption that without external field the vacuum is given by $P^0_-$.  This choice is not physically correct: it corresponds to neglecting the interaction between the virtual electrons. This deficiency was recently overcome by Hainzl, Lewin and Solovej \cite{HLSo} who used a thermodynamic limit applied to the QED Hamiltonian restricted to Hartree-Fock states (mean-field approximation), in order to define the free vacuum. Doing so, they obtained a slightly different translation-invariant projector $\cP^0_-$, solution of a certain self-consistent equation. Then, they showed by the same thermodynamic limit procedure that in the external field case the BDF model should better rely on this new free vacuum instead of Dirac's choice $P^0_-$ used by Chaix and Iracane. Note that the projector $\cP^0_-$ had been constructed earlier by Lieb and Siedentop \cite{LSie}, but the existence proof and the physical interpretation were different.

The Bogoliubov-Dirac-Fock model is a very promising theory: it is well-justified physically, it is better behaved than the usual Dirac-Fock model and it leads to new mathematical problems which are interesting in themselves. In particular, a state of the system always contains infinitely many particles (the real and the virtual ones). This property which raises serious mathematical difficulties is shared with other quantum models, to which a similar study could be applied.

\medskip

The purpose of the present paper is to continue the study which was started in \cite{HLS1,HLS2,HLSo}. In the BDF model, the state of the system is represented by an orthogonal projector of infinite rank 
$$P=\sum_{i\geq 1}|\phi_i\rangle\langle\phi_i|,$$
where $(\phi_i)_{i\geq1}$ is an orthonormal basis of $\text{Ran}(P)$. The projector $P$ should be seen as the one-body density matrix of the following formal wavefunction depending on infinitely many variables
\begin{equation}
\Psi=\phi_1\wedge\phi_2\wedge\cdots\wedge\phi_i\wedge\cdots,
\label{form_psi_intro}
\end{equation}
which is a kind of infinite Hartree-Fock state. Here $\wedge$ denotes the usual wedge product of functions in $L^2(\R^3,\C^4)$.
The projector $P$ represents the whole system consisting of both the real and the virtual particles of the Dirac sea, but there is no distinction between them \emph{a priori}. It is only for the solution of the problem that the real particles will be identified and separated from the virtual ones. The BDF energy $\E^\nu$ is a nonlinear functional of the variable $P$, which is bounded below on an appropriate set which will be defined later. The expression of $\E^\nu$ depends on a fixed external charge density $\nu$.

In \cite{HLS1,HLS2}, we proved the existence of a \emph{global minimizer} of the BDF energy, interpreted as the polarized vacuum in the electrostatic field $V=-\alpha\nu\ast|\cdot|^{-1}$ created by the external density $\nu$. Here $\alpha=e^2$ is the bare coupling constant and $-e$ is the bare charge of an electron. The density $\nu$ has no sign \emph{a priori} but when $\nu\geq0$, one may think of it as the density of a system of nuclei. In this paper we study the minimization of $\E^\nu$ under the constraint that the charge of our state $P$ is equal to $-eN$ where $N$ is some integer. Of course the total charge of a state of the form \eqref{form_psi_intro} is formally infinite since $\Psi$ represents infinitely many negatively charged particles. But one can define the difference between the charge of $P$ and the (infinite) charge of the free vacuum. It is this difference which is fixed to $-eN$. A precise definition will be given below.

This \emph{charge constrained minimization problem} is much more delicate than the global minimization of \cite{HLS1,HLS2}. When $\nu\geq0$ and $N>0$, a minimizer in the $-eN$ charge sector will usually represent the state of $N$ electrons interacting with the polarized vacuum and the external field $V=-\alpha\nu\ast|\cdot|^{-1}$. As usual for Hartree-Fock type theories \cite{LS,Lieb2,Lions-87}, one does not expect that this minimizer will always exist. Indeed, if $\nu$ is not strong enough to bind the $N$ electrons together with the polarized vacuum, there should be no minimizer. On the other hand, if $\nu$ is too strong, some electron-positron pairs could be created.

Let us denote by $E^\nu(N)$ the infimum of the BDF energy $\E^\nu$ in the charge sector $-eN$ and in the presence of the external density $\nu$ (a precise definition will be given in Section \ref{min_charge_sector}). Our main result (Theorem \ref{HVZ}) will be the statement that all the minimizing sequences for $E^\nu(N)$ are precompact if and only if a HVZ-type inequality holds:
\begin{equation}
\forall K\in\Z\setminus\{0\},\quad E^\nu(N)<E^\nu(N-K)+E^0(K).
\label{HVZ_intro}
\end{equation}
Inequalities like \eqref{HVZ_intro} are very common in the study of linear \cite{H,V,Z,GLL} and nonlinear \cite{Lions-84,Lions-87} systems. Assume $N>0$ for simplicity. Then $E^\nu(N)$ is the infimum of the energy of a system of $N$ electrons coupled to the Dirac sea. When $0<K\leq N$, \eqref{HVZ_intro} means that it is not favorable to let $K$ electrons escape to infinity, while keeping $N-K$ electrons near the nuclei. When $K<0$, it means that it is not favorable to let $\vert K\vert$ positrons escape to infinity, while keeping $N+\vert K\vert$ electrons near the nuclei. When $K>N$, it means that it is not favorable to let $K$ electrons escape to infinity, while keeping $K-N$ positrons near the nuclei. 
When $\alpha$ is small enough and $N>0$, it will be shown that the separation of electron-positron pairs is not energetically favorable, so that one just needs to check \eqref{HVZ_intro} for $K=1,2,...,N$.

From a mathematical point of view, proving the compactness of minimizing sequences assuming \eqref{HVZ_intro} is a subtle task. Indeed, the fact that our main variable $P$ is a projector of infinite rank complicates a lot the study of minimizing sequences, for instance compared to the Hartree-Fock case \cite{LS,Lions-87} in which only finitely many particles are described. In particular, it is not obvious at all to localize our state $P$ in space in order to decouple the electrons staying close to the nuclei from those which escape to infinity. These complications are consequences of the vacuum effects. Similar issues are encountered in the study of other models describing systems of infinitely many particles. It is our hope that this work will provide a better understanding of these other models too.

When \eqref{HVZ_intro} holds, there exists a minimizer in the charge sector $-eN$. This is an orthogonal projector $P$ satisfying some nonlinear equation of the form
\begin{equation}
P=\chi_{(-\ii,\mu]}\left(D^0-\alpha\nu\ast|\cdot|^{-1}+V_P\right)
\label{equation_intro}
\end{equation}
where $V_P$ is an operator depending on $P$ itself and $\mu$ is a Lagrange multiplier associated with the charge constraint, interpreted as a chemical potential. The same equation was obtained in \cite{HLS1,HLS2,HLSo} for the vacuum case (global minimization), but with $\mu=0$. When the external density $\nu$ is not too strong and $N>0$, then it will hold $\mu>0$ and the operator $D^0-\alpha\nu\ast|\cdot|^{-1}+V_P$ will have exactly $N$ eigenvalues (counted with their multiplicity) in $(0,\mu]$. In this case $P$ can be written
\begin{eqnarray}
P & = & \chi_{(-\ii,0]}\left(D^0-\alpha\nu\ast|\cdot|^{-1}+V_P\right)+\chi_{(0,\mu]}\left(D^0-\alpha\nu\ast|\cdot|^{-1}+V_P\right)\nonumber\\
 & := & P_{\rm vac}+\sum_{n=1}^N|\phi_n\rangle\langle\phi_n|.
\label{decom_intro}
\end{eqnarray}
Formula \eqref{decom_intro} allows to distinguish the ``real" electrons (represented by the orbitals $(\phi_n)_{n=1}^N$) from the self-consistent polarized vacuum $P_{\rm vac}$. As explained in Section \ref{min_charge_sector}, the orbitals $(\phi_1,...,\phi_N)$ are solutions of a Dirac-Fock type system of equations \cite{ES1}, in which the mean-field operator is perturbed by the self-consistent vacuum polarization potentials.

In the present work, we shall provide two regimes in which the condition \eqref{HVZ_intro} holds, and therefore for which a BDF minimizer exists in the charge sector $-eN$. The first is the \emph{weak coupling regime} in which the coupling constant $\alpha\ll1$, but $\alpha\nu$ (hence $\alpha Z$) and $N$ are both fixed (Theorem \ref{exists_weak_limit}). The second is the \emph{nonrelativistic regime} $c\gg1$ with $\nu$ and $N$ fixed (Theorem \ref{non-rel}). In the latter case, we also prove that the $N$ orbitals $(\phi_1,...,\phi_N)$ of \eqref{decom_intro} converge to a ground state of the nonrelativistic Hartree-Fock functional \cite{LS,Lions-87} as $c\to\ii$. A similar result was already obtained by Esteban and Séré in the Dirac-Fock case \cite{ES2}.

\medskip

The paper in organized as follows. In the first section, we define properly the model and state our main results. Section \ref{prelim} is devoted to the proof of some preliminary results which will be needed throughout the paper. The last three sections are then devoted to the proofs of Theorems \ref{HVZ}, \ref{exists_weak_limit} and \ref{non-rel}.

\bigskip

\noindent\textbf{Acknowledgement.}
The authors acknowledge support from the European Union's IHP network \emph{Analysis \& Quantum} HPRN-CT-2002-00277. M.L. and E.S. acknowledge support from the project \emph{``ACCQUAREL"} NT05-4\_44652 funded by the French National Research Agency GIP-ANR.

\section{Model and main results}
\subsection{The mean-field approximation in no-photon QED}
We start by recalling briefly the physical meaning of the model, as explained in \cite{HLSo}. As mentioned in the introduction, the state of our system is represented by an infinite-rank orthogonal projector $P$, which is seen as the density matrix of an infinite Hartree-Fock state \eqref{form_psi_intro}. We recall that although $P$ should be interpreted as the state of the coupled system `real particles + vacuum', there is no canonical distinction between the real and virtual particles \emph{a priori}. For $N$ electrons, this would correspond to a decomposition of the form $P=P_{\rm vac} + \gamma$ where $\gamma$ is an orthogonal projector of rank $N$ satisfying $P_{\rm vac}\gamma=\gamma P_{\rm vac}=0$ (for $N$ positrons, this becomes $P=P_{\rm vac} - \gamma$). There are infinitely many such decompositions for a given $P$ and a given $N$. But for the solution of our equation, a particular decomposition may be chosen in a natural way.

The energy of the system in the Hartree-Fock state $P$ can be deduced from the QED Hamiltonian formalism in Coulomb gauge and neglecting photons, see \cite{HLSo}. The energy functional is \emph{formally}
\begin{equation}\label{def_energy_QED}
P\mapsto \mathcal{E}_{\rm QED}^\nu(P-1/2)
\end{equation}
where 
\begin{multline}\label{formal_energy}
\mathcal{E}^\nu_{\rm QED}(\Gamma):= \tr(D^0\Gamma) + \int_{\R^3}V(x)\rho_\Gamma(x)\,dx \\
+\frac{\alpha}{2}\iint_{\R^3\times\R^3}\frac{\rho_\Gamma(x)\rho_\Gamma(y)}{|x-y|}dx\,dy
- \frac{\alpha}{2}\iint_{\R^3\times\R^3}\frac{|\Gamma(x,y)|^2}{|x-y|}dx\,dy,
\end{multline}
$V=-\alpha\nu\ast|\cdot|^{-1}$ being the external electrostatic potential created by the density of charge $\nu$.
If $\nu\geq0$ it can for instance represent a system of nuclei in a molecule, but in most of the results of this paper the sign of $\nu$ needs not be fixed. Also we shall not allow pointwise nuclei and $\nu$ will essentially be an $L^1_{\rm loc}$ function. It is a well known phenomenon in QED that pointwise nuclei create spurious divergences (see e.g., \cite{K}). But the regularity assumption on $\nu$ is not really a restriction from the point of view of physics: point-like nuclei do not exist in nature.
In \eqref{formal_energy}, $\rho_\Gamma$ is the density defined formally as $\rho_\Gamma(x)=\tr_{\C^4}(\Gamma(x,x))$, and $\alpha$ is the \emph{bare coupling constant}, $\alpha=e^2$. 

In QED, a global minimizer of $P\mapsto \mathcal{E}_{\rm QED}^\nu(P-1/2)$ represents the vacuum, whereas other types of states (for $N$ electrons for instance) are obtained by minimizing this functional with a charge constraint. The charge is formally defined as 
\begin{equation}
-e\int_{\R^3}\rho_{[P-1/2]}(x)\,dx\; ``="\; (-e)\tr(P-1/2).
\label{def_charge}
\end{equation}

The subtraction of half the identity in \eqref{def_energy_QED} and \eqref{def_charge} is a kind of \emph{renormalization} which was introduced by Heisenberg \cite{Hei} and has been widely used by Schwinger (see \cite[Eq. $(1.14)$]{Sch1}, \cite[Eq. $(1.69)$]{Sch2} and \cite[Eq. $(2.3)$]{Sch3}) as a necessity for a covariant formulation of QED.

Of course, the expression of the mean-field QED energy \eqref{formal_energy} is purely formal: if $P$ is an orthogonal projector in infinite dimension, $P-1/2$ is never compact and therefore $\mathcal{E}_{\rm QED}^\nu(P-1/2)$ is not well defined. Even the density of charge $\rho_{[P-1/2]}$ is not a well defined object. For this reason, it was proposed in \cite{HLSo} to use a thermodynamic limit in order to give a rigorous meaning to the minimization of \eqref{formal_energy}: the idea is to define the model in a bounded domain in space, and a cut-off $\Lambda$ in Fourier space. This was done in \cite{HLSo} in a box $\cC_L=[-L/2;L/2)^3$ with periodic boundary conditions, and cutting the Fourier expansion outside a ball of radius $\Lambda$. Then, the minimization in $\cC_L$ makes perfectly sense ($L^2(\R^3,\C^4)$ has been replaced by a finite-dimensional space), and one can study the limit of the sequence of minimizers when $L\to\ii$.

In \cite{HLSo}, this technique was used to define properly the free vacuum and justify the validity of the BDF functional. Notice that the ultraviolet cut-off $\Lambda$ is fixed and will \emph{not} be removed: it is well-known that QED contains problematic ultraviolet divergences which are difficult to deal with. We therefore introduce the following functional space
$$\gH_\Lambda:=\left\{f\in L^2(\R^3,\C^4),\ {\rm supp}(\widehat{f})\subset B(0,\Lambda)\right\}.$$
Notice that $\gH_\Lambda$ is contained in the domain $H^1(\R^3,\C^4)$ of $D^0$, and that $D^0\gH_\Lambda=\gH_\Lambda$. In the following, we still denote by $D^0$ its restriction to $\gH_\Lambda$.
Taking $\nu=0$ in \eqref{formal_energy} (free case) and studying the thermodynamic limit $L\to\ii$, the free vacuum was obtained in \cite{HLSo}. It is a \emph{translation-invariant} projector $\cP^0_-$ satisfying the Euler-Lagrange equation
\begin{equation}
\left\{\begin{array}{l}
\displaystyle \cP^0_-=\chi_{(-\ii,0)}\left(\cD^0\right),\\
\displaystyle\cD^0=D^0-\alpha\frac{(\cP^0_- - 1/2)(x,y)}{|x-y|}.
\end{array} \right.
\label{scf_proj}
\end{equation}
The operator $\cD^0$ which appears in \eqref{scf_proj} is a translation-invariant operator taking the following special form \cite{LSie,HLSo}, in the Fourier space,
\begin{equation}
\cD^0(p)=\alp\cdot \omega_p\, g_1(|p|)+g_0(|p|)\beta,\qquad \omega_p=p/|p|.
\label{form_nouveau_D0}
\end{equation}
Here $g_1$ and $g_0$ are real and smooth functions satisfying 
\begin{equation}
x\leq g_1(x)\leq x\,g_0(x).
\label{prop_g0_g1}
\end{equation}
Note the self-consistent equation \eqref{scf_proj} was already solved by Lieb and Siedentop \cite{LSie}, but their interpretation was not variational. They used a fixed point approach valid when $\alpha\log\Lambda< C$. In \cite{HLSo}, the free vacuum $\cP^0_-$ solution of \eqref{scf_proj} is constructed as a minimizer of the energy per unit volume for any value of the ultraviolet cut-off $\Lambda$, and under the condition $0\leq\alpha<4/\pi$. This last inequality is related to Kato's inequality $|x|^{-1}\leq \pi/2|\nabla|$. Hence, in the whole paper we shall assume that $0\leq\alpha<4/\pi$. We use the following notation $\cP^0_+=1-\cP^0_-$.

The mean-field approximation in no-photon QED is therefore very close to the original Dirac's picture of the free vacuum, the latter being described as an infinite rank spectral projector associated with the negative spectrum of a translation-invariant Dirac-type operator. However, it does not correspond exactly to the original ideas of Dirac: when $\alpha\neq0$, $\cP^0_-$ is different from $P^0_-$. Even the free vacuum $\cP^0_-$ is solution of a complicated nonlinear equation \eqref{scf_proj}. This is because the interaction between the virtual particles is taken into account, similarly to the real ones. The Dirac picture is only recovered in the non-interacting case $\alpha=0$. 

It is important physically that the so-obtained free vacuum is invariant by translations. This means that the density of charge $\rho_{[\cP^0_--1/2]}$ is (formally) constant. More precisely, the subtraction of half the identity allows to obtain a vanishing density, $\rho_{[\cP^0_--1/2]}\equiv0$. By \eqref{form_nouveau_D0}, we have
$$\cP^0_-(p)-1/2=-\frac{g_1(|p|)\alp\cdot \omega_p +g_0(|p|)\beta}{2\sqrt{g_1(|p|)^2+g_0(|p|)^2}},$$
from which we infer that $\tr_{\C^4}[(\cP^0_--1/2)(p)]=0$ for any $p\in B(0,\Lambda)$, the Pauli matrices being trace-less. Thus the (constant) density of charge of the free vacuum vanishes:
$$\rho_{[\cP^0_--1/2]}\equiv (2\pi)^{-3}\int_{B(0,\Lambda)}\tr_{\C^4}(\cP_-^0(p)-1/2)\, dp=0.$$
This formally means that
\begin{equation}
\text{``}\,\tr\left(\cP^0_- - 1/2\right)=\int_{\R^3}\rho_{[\cP^0_--1/2]}(x)\, dx=0\,\text{''}
\label{charge_nulle}
\end{equation}
and therefore that the free vacuum is not charged.

As a consequence of \eqref{form_nouveau_D0}, the spectrum of $\cD^0$ is
$$\sigma(\cD^0)=\left\{\pm \sqrt{g_0(|p|)^2+g_1(|p|)^2},\ p\in B(0,\Lambda)\right\}.$$
It has a gap which is greater than the one of $D^0$, by \eqref{prop_g0_g1}:
\begin{equation}
1\leq m(\alpha):=\min\sigma(|\cD^0|).
\label{def_threshold}
\end{equation}
In Lemma \ref{threshold} below, we shall prove that when $\alpha\ll1$, then $m(\alpha)=g_0(0)$. We conjecture that this is true for any $0\leq \alpha<4/\pi$. Notice that the following expansion is known \cite{LSie,HLSo}: $g_0(0)=1+\frac{\alpha}{\pi}\rm{arcsinh}(\Lambda)+O(\alpha^2)$.

\medskip

Once the free vacuum is defined, in the external field case $\nu\neq0$ one can measure the energy of any state $P$ with respect to the (infinite) energy of the free vacuum $\cP^0_-$. The Bogoliubov-Dirac-Fock energy is formally defined as
\begin{eqnarray}
``\,\E^\nu(P-\cP^0_-) & = & \E_{\rm QED}^\nu(P-1/2)-\E_{\rm QED}^0(\cP^0_--1/2)\label{formal_BDF2}\\
 & = & \tr\left(\cD^0Q\right) + \int_{\R^3}\!\!V(x)\rho_{Q}(x)dx+\frac{\alpha}{2}\iint_{\R^3\times\R^3}\!\!\!\!\!\frac{\rho_{Q}(x)\rho_Q(y)}{|x-y|}dx\,dy\nonumber\\
  & & \qquad - \frac{\alpha}{2}\iint_{\R^3\times\R^3}\frac{|Q(x,y)|^2}{|x-y|}dx\,dy\,",\nonumber
\end{eqnarray}
where we have introduced $Q=P-\cP^0_-=(P-1/2)-(\cP^0_--1/2)$ and used the definition of the self-consistent free Dirac operator $\cD^0$, see \eqref{form_nouveau_D0}. This functional has been mathematically defined and studied in \cite{HLS1,HLS2}, but with $P^0_-$ as reference. In \cite{HLSo}, it was proved that any sequence of global minimizers of the full QED energy in a finite box of size $L$ converges (up to a subsequence) to a global minimizer of the BDF functional as $L\to\ii$, justifying the formal derivation \eqref{formal_BDF2}. Notice that by \eqref{charge_nulle}, the total charge of our state is now formally given by
$$``-e\tr(P-1/2)=-e\tr(Q)=-e\tr(P-\cP^0_-)".$$
In the next section, we define properly the BDF functional $\E^\nu$ and recall its main properties proved in \cite{HLS1,HLS2,HLSo}. It was noticed in \cite{HLS1} that the minimizer of the BDF energy cannot be searched in the trace-class. For this reason, it was necessary to extend the definition of the trace in order to give a meaning to $\tr(P-\cP^0_-)$ and to the energy \eqref{formal_BDF2}. In this paper, we use all this formalism.

\subsection{The Bogoliubov-Dirac-Fock theory}
We denote by $\gS_p(\gH)$ the usual Schatten class of compact operators $A$ acting on a Hilbert space $\gH$ and such that $\tr(|A|^p)<\ii$, see, e.g., \cite{RS1}, and by $\cB(\gH)$ the space of bounded operators on $\gH$.
We recall \cite{HLS1} that a Hilbert-Schmidt operator $A\in\gS_2(\gH_\Lambda)$ is said to be $\cP^0_-$--trace class if $A^{++}=\cP^0_+A\cP^0_+$ and $A^{--}=\cP^0_-A\cP^0_-$ both belong to the trace-class $\gS_1(\gH_\Lambda)$ (but $A^{+-}=\cP^0_+A\cP^0_-$ and $A^{-+}=\cP^0_-A\cP^0_+$ need only be Hilbert-Schmidt). We denote by $\gS_1^{\cP^0_-}(\gH_\Lambda)$ this subspace of $\gS_2(\gH_\Lambda)$. We define the $\cP^0_-$--trace of $A$ as
$$\tr_{\cP^0_-}(A)=\tr(A^{++})+\tr(A^{--}).$$
We refer to \cite[Section 2.1]{HLS1} for a general definition valid for any reference projector and for the useful properties which will be needed in this paper.
The BDF energy reads \cite{HLS1,HLS2,HLSo} 
\begin{equation}
\label{BDF1}
\E^\nu(Q):= \tr_{\cP^0_-}(\cD^0Q)-\alpha D(\rho_Q,\nu)+\frac{\alpha}{2}D(\rho_Q,\rho_Q)-\frac{\alpha}{2}\iint_{\R^6}\frac{|Q(x,y)|^2}{|x-y|}dx\,dy
\end{equation} 
where $\nu$ is the smooth density of charge of a system of extended nuclei,
$$D(f,g)=4\pi\int\frac{\overline{\widehat{f}(k)}\widehat{g}(k)}{|k|^2}dk$$
and 
$\cP^0_-$ is the free vacuum defined above.
We define the BDF energy $\E^\nu$ on the convex set
\begin{equation}
\mathcal{Q}_{\Lambda}:=\left\{Q\in\gS^{\cP^0_-}_1(\gH_\Lambda)\ |\ Q^*=Q,\ -\cP^0_-\leq Q\leq \cP^0_+\right\}.
\label{def_Q_Lambda}
\end{equation} 
Notice that $\mathcal{Q}_{\Lambda}$ is the closed convex hull of the set of operators of the form $P-\cP^0_-\in\gS_2(\gH_\Lambda)$ where $P$ is an orthogonal projector \cite[Lemma 2]{HLS1}. Studying the BDF energy on $\cQ_\Lambda$ will be easier and minimizers will be shown to be extremal points, i.e. of the form $Q=P-\cP^0_-$. This is a very common technique for Hartree-Fock theories \cite{Lieb}.

\begin{remark}\rm 
Notice that compared to \cite{CI,CIL,Chaix,BBHS,HLS1,HLS2}, we have not only replaced $P^0_-$ by $\cP^0_-$, but also $D^0$ by $\cD^0$ in the definition \eqref{BDF1} of the BDF energy, following the results of \cite{HLSo}.
\end{remark}

Any $Q\in\cQ_\Lambda\subset\gS_2(\gH_\Lambda)$ has a well-defined integral kernel denoted by $Q(x,y)$, such that its Fourier transform $\widehat{Q}(p,q)$ is supported in $B(0,\Lambda)\times B(0,\Lambda)$. Therefore the function $Q(x,y)$ appearing in \eqref{BDF1} is smooth and the charge density $\rho_Q(x):=\tr_{\C^4}Q(x,x)$ is also a well-defined object \cite{HLS1}. In Fourier space, 
\begin{equation}
\widehat{\rho_Q}(k)=(2\pi)^{-3/2}\int_{\substack{|p+k/2|\leq\Lambda\\ |p-k/2|\leq\Lambda}}\tr_{\C^4}\left(\widehat{Q}(p+k/2,p-k/2)\right)\, dp,
\label{def_rho}
\end{equation}
which shows that $\rho_Q\in L^2(\R^3)$.
Introducing the so-called Coulomb space $\mathcal{C}=\left\{f\ |\ D(f,f)<\ii\right\}$, the linear map $Q\in\gS_1^{\cP^0_-}(\gH_\Lambda) \longmapsto \rho_Q\in\cC\cap L^2(\R^3,\R)$ 
is continuous when $\gS_1^{\cP^0_-}(\gH_\Lambda)$ is equipped with the Banach space norm
$$\norm{Q}_{1;\cP^0_-}:=\norm{Q^{++}}_{\gS_1(\gH_\Lambda)}+\norm{Q^{--}}_{\gS_1(\gH_\Lambda)}+\norm{Q^{+-}}_{\gS_2(\gH_\Lambda)}+\norm{Q^{-+}}_{\gS_2(\gH_\Lambda)},$$
as shown in the following useful result, proved in Appendix A.
\begin{lemma}[Continuity of the map $Q\mapsto\rho_Q$]\label{lemma_continuity} Assume that $0\leq\alpha_0<4/\pi$ and $\Lambda>0$. Then there exists a constant $C_{\Lambda,\alpha_0}$ such that
\begin{equation}
\label{continuity}
\forall0\leq\alpha\leq\alpha_0,\ \forall Q\in\gS_1^{\cP^0_-}(\gH_\Lambda),\qquad \norm{\rho_Q}_{L^2}+D(\rho_Q,\rho_Q)^{1/2}\leq C_{\Lambda,\alpha_0} \norm{Q}_{1;\cP^0_-}.
\end{equation}
\end{lemma}

It has been proved in \cite{HLS1} that $\E^\nu$ is well-defined and bounded-below on $\cQ_\Lambda$, independently of $\Lambda$ (see also \cite[Theorem 1]{HLS2} and \cite[Theorem 2.5]{HLSo}):
\begin{equation}
 \forall Q\in\mathcal{Q}_\Lambda,\qquad \E^\nu(Q)+\frac\alpha2 D(\nu,\nu)\geq0.
\label{bound_below}
\end{equation}
The proof of \eqref{bound_below} was itself essentially contained in \cite{BBHS}. If moreover $\nu=0$, then $\E^0$ is non-negative on $\mathcal{Q}_\Lambda$, i.e. we recover that $0$ is its unique minimizer.

We shall need to endow $\cQ_\Lambda$ with a weak topology for which the unit ball is compact. We recall that $\gS_1(\gH_\Lambda)$ is the dual of the space of compact operators \cite[Thm VI.26]{RS1}. It can therefore be endowed with the associated weak-$\ast$ topology. This allows to define a weak topology on $\gS_1^{\cP^0_-}(\gH_\Lambda)$ for which $Q_n\wto Q$ means $Q_n\wto Q$ in $\gS_2(\gH_\Lambda)$, 
$$\lim_{n\to\ii}\tr(Q^{++}_nK)=\tr(Q^{++}K)\quad \text{and}\quad \lim_{n\to\ii}\tr(Q^{--}_nK)=\tr(Q^{--}K)$$
 for any compact operator $K$. It was proved\footnote{In \cite{HLS2}, the BDF energy is studied on a set ${\mathcal S}_\Lambda$ slightly different from $\cQ_\Lambda$ but the arguments used to prove \cite[Thm 1]{HLS2} can be adapted by means of Lemma \ref{lemma_continuity}.} in \cite[p. 4492]{HLS2} that $\E^\nu$ is weakly lower semi-continuous (wlsc) for this topology on the convex set $\cQ_\Lambda$, and it therefore possesses a global minimizer $\bar Q=\cP_--\cP^0_-\in\cQ_\Lambda$ where $\cP_-$ is an orthogonal projector satisfying the equation
\begin{equation}
\cP_-  =  \chi_{(-\infty, 0]}\left(\cD_{\bar Q}\right)=\chi_{(-\infty, 0]}\left( \cD^0+\alpha\left(\rho_{\bar Q}-\nu\right)\ast\frac{1}{|\cdot|}-\alpha\frac{\bar Q(x,y)}{|x-y|}\right)
\label{scf_Q_vide1}
\end{equation}
Additionally, if $\alpha D(\nu,\nu)^{1/2}$ is small enough \cite[Eq. $(11)$]{HLS2}, this minimizer $\bar Q$ is unique and the charge of the polarized vacuum vanishes: $\tr_{\cP^0_-}(\bar Q)=0$.

\subsection{Existence of atoms and molecules}\label{results}
\subsubsection{Minimization of $\E^\nu$ in charge sectors}\label{min_charge_sector}
We consider the following variational problem
\begin{equation}
\label{def_min}
E^\nu(q)=\inf_{Q\in\mathcal{Q}_\Lambda(q)}\E^\nu(Q)
\end{equation}
where $\mathcal{Q}_\Lambda(q)$, the sector of charge $-eq$, is defined as
$$\cQ_\Lambda(q):=\{Q\in\mathcal{Q}_\Lambda,\ \tr_{\cP^0_-}(Q)=q\}$$
and $q$ is any real number.

As recalled before, it is known that the polarized vacuum (i.e. the global minimizer of $\E^\nu$) is a solution of $E^\nu(0)$ when $\alpha D(\nu,\nu)^{1/2}$ is small enough. But in general, it is not obvious at all to prove the existence of a solution to \eqref{def_min}. This is because even if the energy functional is wlsc, the charge sectors $\cQ_\Lambda(q)$ are not closed for the weak topology of $\cQ_\Lambda$: a weakly converging sequence might loose or gain some charge.
We now describe the properties of a minimizer of \eqref{def_min} if it exists. We recall that $m(\alpha)$ defined in \eqref{def_threshold} is the threshold of the free mean-field operator $\cD^0$.

\begin{proposition}[Self-consistent equation solved by a minimizer]\label{prop_scf}
Let be $0\leq\alpha<4/\pi$, $\nu\in\cC$ and $q\in\R$. Any minimizer $Q$, solution of the variational problem \eqref{def_min}, takes the form
$Q=P-\cP^0_--\delta|\phi\rangle\langle\phi|$, where
\begin{equation}
P=\chi_{(-\ii,\mu]}(\cD_Q)=\chi_{(-\ii,\mu]}\left(\cD^0+\alpha(\rho_Q-\nu)\ast1/|\cdot|-\alpha\frac{Q(x,y)}{|x-y|}\right)
\label{scf_eq_min_BDF}
\end{equation}
for some $\mu\in[-m(\alpha),m(\alpha)]$ and where
\begin{enumerate}
\item if $q$ is an integer, then $\delta=0$;
\item if $q$ is not an integer, then $\delta=[q]+1-q$ and $\phi$ is a normalized function of $\ker\left(\cD_Q-\mu\right)$.
\end{enumerate}
\end{proposition}

The Fermi level $\mu$ is a Lagrange multiplier associated with the charge constraint and interpreted as a chemical potential. The proof of Proposition \ref{prop_scf} is left to the reader. It is an adaptation of proofs in \cite{Bach,BLS,BBHS,HLS1,HLS2} and of the arguments that will be given below for the proof of our other results (see in particular Proposition \ref{lieb}). Notice that when $q=N$ is an integer, then \eqref{scf_eq_min_BDF} means that the last level $\mu$ is necessarily totally filled. This is a general fact for Hartree-Fock type theories \cite{BLLS}.

Equation \eqref{scf_eq_min_BDF} is well known in physics. See \cite[Eq. $(4)$]{RGA} which is exactly equivalent to \eqref{scf_eq_min_BDF} and \cite{RG,ED,GT,DH,Hamm} for related studies. 

Let us assume for simplicity that $q=N$ is an integer. For a minimizer of the form \eqref{scf_eq_min_BDF} and when $N,\mu>0$, it is natural to consider the decomposition $P=P_{\rm vac}+\chi_{(0\,, \,\mu]}(\cD_Q),$
where $P_{\rm vac}$ is the polarized Dirac sea: $P_{\rm vac}:= \chi_{(-\infty\,, \,0]}(\cD_Q)$.
For not too strong external potentials, the vacuum will be neutral,
$\tr_{\cP^0_-}(P_{\rm vac}-\cP^0_-)=0$ 
and therefore $\chi_{(0\,, \,\mu]}(\cD_Q)$ will be a projector of rank $N$:
$$\chi_{(0, \mu]}(\cD_Q)=\sum_{n=1}^N|\phi_n\rangle\langle\phi_n|:=\gamma_\Phi.$$
Then $ \cD_Q\phi_n=\epsilon_n\phi_n,$
where $\epsilon_1\leq\cdots\leq\epsilon_N$ are the $N$ first positive eigenvalues of $\cD_Q$ counted with their multiplicity. Notice that
\begin{multline}
\label{dec_mean_field}
\cD_Q=D^0+\alpha(\rho_\Phi-\nu)\ast\frac{1}{|\cdot|}-\alpha\frac{\gamma_\Phi(x,y)}{|x-y|}\\
+\alpha\rho_{[P_{\rm vac}-1/2]}\ast\frac{1}{|\cdot|}-\alpha\frac{(P_{\rm vac}-1/2)(x,y)}{|x-y|},
\end{multline}
where $\rho_\Phi(x):=\tr_{\C^4}(\gamma_\Phi(x,x))=\sum_{n=1}^N|\phi_n(x)|^2$.
In the first line of \eqref{dec_mean_field}, the Dirac-Fock operator associated with $(\phi_1,...,\phi_N)$ appears, see \cite{ES1}. This shows that the electronic orbitals $\phi_i$ are solutions of a Dirac-Fock type equation in which the mean-field operator $\cD_Q$ is perturbed by the (self-consistent) potential of the Dirac sea $P_{\rm vac}-1/2$. Of course, the totally new feature is that these equations have been obtained by a minimization principle (as first proposed in \cite{CI}) while in the Dirac-Fock theory the energy functional is not bounded from below. The Dirac-Fock model is thus seen as a \emph{non-variational} approximation of the mean-field model of no-photon QED: the Euler-Lagrange equations are similar but the variational structure is very different.
In Theorem \ref{non-rel} below, we shall prove that the orbitals $(\phi_1,...,\phi_N)$ converge to a Hartree-Fock ground state \cite{LS,Lions-87} in the non-relativistic limit.

If $N,\mu<0$ a similar decomposition can be applied,
$$P=P_{\rm vac}-\chi_{(\mu\,, \,0)}(\cD_Q).$$ When the polarized vacuum is neutral, $\tr_{\cP^0_-}(P_{\rm vac}-\cP^0_-)=0$, we obtain
$$P=P_{\rm vac}-\sum_{n=1}^N|\phi_n\rangle\langle\phi_n|$$
where the minus sign reflects that the orbitals $(\phi_n)_{n=1}^N$ describe positrons (up to charge conjugation). It holds 
$\cD_Q\phi_n=\epsilon_{-n}\phi_n$
where $\epsilon_{-1}\geq\cdots\geq\epsilon_{-N}$ are the $N$ highest negative eigenvalues of $\cD_Q$ counted with their multiplicity. The multiplier $\mu$ is chosen to ensure $\epsilon_{-N}>\mu\geq\epsilon_{-N-1}$.

\begin{remark}\rm
To any electronic solution with density $\nu$, one can associate a positronic solution with density $-\nu$ by charge conjugation \cite[Remark 8]{HLS1}.
\end{remark}

\subsubsection{A dissociation criterion}
The main result of this paper is the following
\begin{theorem}[Binding Conditions \& Existence of a Ground State]\label{HVZ} Let be $0\leq\alpha< 4/\pi$, $\Lambda>0$, $\nu\in\cC$ and $q\in\R$. Then the following assertions are equivalent
\begin{description}
\item[$(H_1)$] for any $k\in\R\setminus\{0\}$, $E^\nu(q) < E^\nu(q-k)+E^0(k)$;
\item[$(H_2)$] each minimizing sequence $(Q_n)_{n\geq1}$ for $E^\nu(q)$ is precompact in $\cQ_\Lambda$ and converges, up to a subsequence, to a minimizer $Q$ of $E^\nu(q)$. 
\end{description}
If moreover $q=N\in\Z$ is an integer, then $(H_1)$ can be replaced by
\begin{description}
\item[$(H_1')$] for any $K\in\Z\setminus\{0\}$, $E^\nu(N) < E^\nu(N-K)+E^0(K)$.
\end{description}
When $(H_2)$ holds true, the operator $Q$ is a solution of the self-consistent equation \eqref{scf_eq_min_BDF}
for some Lagrange multiplier $\mu\in[-m(\alpha),m(\alpha)]$.
\end{theorem}

\begin{remark}\rm
Notice that the inequality $E^\nu(q) \leq E^\nu(q-k)+E^0(k)$ is true for any $q\in\R$ and any $k\in\R$, as proved later in Proposition \ref{HVZ_large}.
\end{remark}
\begin{remark}\rm
It will be proved in Lemma \ref{encadrement_min_prop} below that $\lim_{|q|\to\ii}E^\nu(q)=\ii$.
This implies that for any fixed $q$, there exists a constant $M$ such that $|k|\geq M\Longrightarrow E^\nu(q)<E^\nu(q-k)+E^0(k)$.
When $q=N>0$ is a positive integer and $\alpha$ is small enough (see  Corollary \ref{simplification_HVZ} below for a precise estimate)
then it holds $E^\nu(N)<E^\nu(N-K)+E^0(K)$ for all $K>N$ and $K<0$.
In this case, $(H_1')$ can be replaced by the more usual condition
\begin{description}
\item[$(H_1'')$]  $E^\nu(N)<\min\{E^\nu(N-K)+E^0(K),\ K=1,...,N\}$.
\end{description}
\end{remark}

Conditions like $(H_1)$ appear classically when analyzing the compactness properties of minimizing sequences, for instance by using the concentration-compactness principle of P.-L. Lions \cite{Lions-84,Lions-87}. They are also very classical for $N$-body Hamiltonians in which the bottom of the essential spectrum has the form of the minimum in the r.h.s. of $(H_1)$, as expressed by the HVZ Theorem \cite{H,V,Z}. In nonrelativistic Quantum Electrodynamics, such binding conditions have also been proved by Griesemer, Lieb and Loss \cite{GLL,LieLo}.
Notice however that, unlike usual HVZ-type results in which a condition similar to $(H_1)$ appears only for $0<k\leq N$, here one has to verify that these strict inequalities hold for any $k\neq 0$. The reason is that electron-positron pairs can appear.

For the sake of simplicity, we assume in the following that $q=N$ is an integer. In the next two sections, we provide two regimes in which $(H_1')$ is true. 

\subsubsection{Existence of a minimizer in the weak coupling regime}
We consider first the weak coupling regime $\alpha\ll1$ and $\bar\nu:=\alpha\nu$ is fixed (the number $N$ of electrons is also fixed). Our result is the following:
\begin{theorem}[Binding Conditions in the weak coupling regime]\label{exists_weak_limit} Assume that the ultraviolet cut-off $\Lambda$ is fixed, and that $\bar\nu\in\cC$ is such that ${\rm ker}(D^0-\bar\nu\ast|\cdot|^{-1})=\{0\}$. Then for any integer $N\in\Z$, one has 
\begin{equation}
\lim_{\alpha\to0}E^{\bar\nu/\alpha}(N)=\inf_{\substack{Q\in\gS^{P^0_-}_1(\gH_\Lambda),\ -P^0_-\leq Q\leq P^0_+,\\ \tr_{P^0_-}Q=N}}\tr_{P^0_-}\left\{ (D^0-\bar\nu\ast|\cdot|^{-1})Q\right\}.
\label{limit_weak_coupling}
\end{equation}

If we moreover assume that $N\geq0$ and that $\bar\nu\in\cC$ is such that
\begin{description}
\item[$(a)$] $\sigma(D^0-\bar\nu\ast|\cdot|^{-1})$ contains at least $N$ positive eigenvalues below $1$,
\item[$(b)$] ${\rm ker}(D^0-t\bar\nu\ast|\cdot|^{-1})=\{0\}$ for any $t\in[0;1]$,
\end{description}
then $(H_1')$ holds in Theorem \ref{HVZ} for $\alpha$ small enough, and therefore there exists a minimizer $Q_\alpha$ of $E^{\bar\nu/\alpha}(N)$. It takes the form
\begin{equation}
Q_\alpha=\chi_{(-\ii,0]}\left(\cD_{Q_\alpha}\right)-\cP^0_-+\chi_{(0,\mu_\alpha]}\left(\cD_{Q_\alpha}\right):=Q_\alpha^{\rm vac}+\sum_{i=1}^N|\phi_i^\alpha\rangle\langle\phi_i^\alpha|,
\label{form_Q_weak_coupling}
\end{equation}
\begin{equation}
\cD_{Q_\alpha}\phi_i^\alpha=\epsilon_i^\alpha\phi_i^\alpha
\label{form_Q_weak_coupling2}
\end{equation}
where $\epsilon_1^\alpha\leq\cdots\leq \epsilon_N^\alpha$ are the $N$ first positive eigenvalues of $\cD_{Q_\alpha}$.
Finally, for any sequence $\alpha_n\to0$, $(\phi_1^{\alpha_n},...,\phi_N^{\alpha_n})$ converges (up to a subsequence) in $\gH_\Lambda$ to $(\phi_1,...,\phi_N)$ which are $N$ first eigenfunctions of $D^0-\bar\nu\ast|\cdot|^{-1}$ and $Q_{\alpha_n}^{\rm vac}$ converges to $\chi_{(-\ii;0)}\left(D^0-\bar\nu\ast|\cdot|^{-1}\right)-P^0_-$ in $\gS_2(\gH_\Lambda)$.
\end{theorem}

Notice that $(b)$ means that no eigenvalue crosses 0 when $t$ is increased from 0 to 1. It is easy to give conditions for which $(a)$ and $(b)$ are satisfied. For instance, one can assume that $\bar\nu\in\cC\cap L^1(\R^3)$, $\bar\nu\geq0$, $\bar\nu\neq0$ and that $\int_{\R^3}\bar\nu\leq 2/(\pi/2+2/\pi)$. This last constant is related to an inequality of Tix \cite{Tix-97,Tix-98}.

\begin{remark}\textnormal{
If $N$ is a negative integer, we are able to prove a similar result if it is assumed instead of $(a)$ that the spectrum $\sigma(D^0-\bar\nu\ast|\cdot|^{-1})$ contains at least $|N|$ \emph{negative} eigenvalues above $-1$.}
\end{remark}

\subsubsection{Existence of a minimizer in the non-relativistic regime}
Next we consider the non-relativistic regime $c\gg1$. For the sake of clarity, we reintroduce the speed of light $c$ in the model and we take $\alpha=1$. The free Dirac operator $D^0$ is then
$$D^0(p)=c\,\alp\cdot p +c^2\beta.$$
The expression of the energy and the definition of the free vacuum $\cP^0_-$ and of the free mean-field operator $\cD^0$ (which of course then depend on $c$ and the ultraviolet cut-off $\Lambda$) are straightforward. To avoid any confusion, we denote by $E^\nu_c(N)$ the minimum energy of the BDF functional. 

In the limit $c\to\infty$, we shall obtain the well-known non-relativistic Hartree-Fock theory \cite{LS,Lions-87}, similarly to the non-relativistic limit of the Dirac-Fock equations studied by Esteban and Séré in \cite{ES2}. For a set of orthonormalized orbitals $\psi=(\psi_1,...,\psi_N)\in (H^1(\R^3,\C^2))^N$, $\int(\psi_i,\psi_j)_{\C^2}=\delta_{ij}$, it reads
\begin{equation}
\E^\nu_{\rm HF}(\gamma_\psi):= \tr((-\Delta/2-\nu\ast|\cdot|^{-1})\gamma_\psi)+\frac12 D(\rho_{\gamma_\psi},\rho_{\gamma_\psi})-\frac12 \iint_{\R^6}\frac{|\gamma_\psi(x,y)|^2}{|x-y|}dx\,dy
\label{HF_energy}
\end{equation}
where $\gamma_\psi=\sum_{i=1}^N|\psi_i\rangle\langle\psi_i|$.
Notice that this model is not posed in $\gH_\Lambda$ but rather in the whole space $H^1(\R^3,\C^2)$ since we shall also be able to remove the ultraviolet cut-off by taking $\Lambda=c\Lambda_0$ for some fixed $\Lambda_0>0$. We define the Hartree-Fock ground state energy as
$$E_{\rm HF}^\nu(N):=\min_{\substack{\psi\in H^1(\R^3,\C^2)^N\\{\rm Gram}_{L^2}\;\psi=Id}}\mathcal{E}^\nu_{\rm HF}(\gamma_\psi).$$
\begin{theorem}[Existence of a minimizer in non-relativistic regime]\label{non-rel} Assume that the ultraviolet cut-off is $\Lambda=c\Lambda_0$ for some fixed $\Lambda_0$. Let be $\nu\in\cC\cap L^1(\R^3,\R^+)$ with $\int_{\R^3}\nu=Z$, and $N$ a positive integer which is such that $Z>N-1$. Then, for $c$ large enough, $(H'_1)$ holds in Theorem \ref{HVZ} and therefore there exists a minimizer $Q_c$ for $E^\nu_{c}(N)$. It takes the following form:
\begin{equation}
Q_c= \chi_{(-\infty,0]}(\cD_{Q_c})-\cP^0_-+\chi_{(0,\mu_c]}(\cD_{Q_c})=Q_c^{\rm vac} +\sum_{i=1}^N|\phi_i^c\rangle\langle\phi_i^c|
\label{eq_nonrel}
\end{equation}
and 
\begin{equation}
\lim_{c\to\infty}\left\{E^\nu_{c}(N)-N\, g_0(0)\right\}= E_{\rm HF}^\nu(N).
\label{limit_energy_nonrel}
\end{equation}
Moreover, for any sequence $c_n\to\infty$, $(\phi_1^{c_n},...,\phi_N^{c_n})$ converges (up to a subsequence) in $H^1(\R^3,\C^4)^N$ towards $\phi=\left(^{\psi}_0\right)$, $\psi\in H^1(\R^3,\C^2)^N$, and where $\gamma_{\psi}$ is a global minimizer of the Hartree-Fock energy \eqref{HF_energy}.
\end{theorem}

It is proved in Lemma \ref{threshold} that $g_0(0)=\min\sigma(|\cD^0|)$ is the threshold of the self-consistent free Dirac operator $\cD^0$ for $c$ large enough.

The rest of the paper is devoted to the proof of Theorems \ref{HVZ}, \ref{exists_weak_limit} and \ref{non-rel}.

%%%%%%%%%%%%%%%%%%%%%%%%%%%%%%%%%%%%%%%%%%%
%%%%%%%%%%%%%%%%%%%%%%%%%%%%%%%%%%%%%%%%%%%
\section{Preliminaries}\label{prelim}
\subsection{Behavior of $E^\nu(q)$ for $|q|\gg1$}
We give conditions which prevent the appearance of electron-positron pairs in minimizing sequences. 
\begin{lemma}\label{encadrement_min_prop} Assume that $0\leq\alpha<4/\pi$, $\Lambda>0$ and $\nu\in\cC$. Then one has
\begin{equation}
\label{encadrement_min}
(1-\alpha\pi/4)m(\alpha)|q|-\frac\alpha2 D(\nu,\nu) \leq E^\nu(q) \leq g_0(0)|q|
\end{equation}
where $g_0(0)\geq1$ is defined in \eqref{form_nouveau_D0} and $m(\alpha)$ is defined in \eqref{def_threshold}.
Hence it holds 
$\lim_{|q|\to\ii}E^\nu(q)=\ii.$
For any fixed $q\in\R$, there exists an $M$ depending on $q$, $\alpha$, $\nu$ such that 
$|k|\geq M\Longrightarrow E^\nu(q)<E^\nu(q-k)+E^0(k)$.
\end{lemma}

\begin{proof}
For the right hand side of \eqref{encadrement_min}, let us fix some orthonormal system $(\psi_1,...,\psi_{[q]+1})$ of smooth $\C^2$-valued functions with compact support in the Fourier domain. We introduce the following 
$$W_\lambda={\rm Span}\left\{\cP^0_+\phi_i^\lambda,\ i=1,...,{[q]+1}\right\},\quad \phi_i^\lambda=\left(\begin{matrix}\psi_i^\lambda\\0\\ \end{matrix}\right)=\left(\begin{matrix}\lambda^{3/2}\psi_i(\lambda\cdot)\\0\\ \end{matrix}\right).$$
Note that for $\lambda$ small enough $W_\lambda$ is a subspace of $\gH_+^0$ of dimension $[q]+1$ since
\begin{equation*}
\pscal{\cP^0_+\phi_i^\lambda,\cP^0_+\phi_j^\lambda} = \pscal{\frac{g_0(|p|)+\sqrt{g_0(|p|)^2+g_1(|p|)^2}}{2\sqrt{g_0(|p|)^2+g_1(|p|)^2}}\psi_i^\lambda,\psi_j^\lambda}
= \delta_{ij}+O(\lambda).
\end{equation*}
Let us choose an orthonormal basis $(\tilde\phi_1^\lambda,...,\tilde\phi_{[q]+1}^\lambda)$ of $W_\lambda$. The r.h.s. of \eqref{encadrement_min} is then obtained by taking a trial state of the form $$Q_\lambda=\epsilon\left(\sum_{i=1}^{[q]}|\tilde\phi^\lambda_i\rangle\langle\tilde\phi^\lambda_i|+(q-[q])|\tilde\phi^\lambda_{[q]+1}\rangle\langle\tilde\phi^\lambda_{[q]+1}|\right)$$
where $\epsilon=1$ if $q>0$, and $\epsilon=-1$ otherwise, and by taking the limit $\lambda\to0$.
To prove the lower bound in \eqref{encadrement_min}, one uses that \cite{BBHS} for any $Q\in\cQ_\Lambda(q)$,
$$\E^\nu_{\rm BDF}(Q)\geq (1-\alpha\pi/4)\tr_{\cP^0_-}(\cD^0Q)-\frac\alpha2 D(\nu,\nu),$$
$$\tr_{\cP^0_-}(\cD^0Q)=\tr(|\cD^0|(Q^{++}-Q^{--}))\geq m(\alpha)\tr(Q^{++}-Q^{--})\geq m(\alpha)|q|.$$
\qed \end{proof}
\begin{corollary}\label{simplification_HVZ} Let be $0\leq\alpha< 4/\pi$, $\nu\in\cC$ and $\Lambda>0$. Assume that $N$ is a non-negative integer and that
\begin{equation}
(g_0(0)-m(\alpha))N+\alpha\left(m(\alpha)(N+2)\frac\pi4+\frac{D(\nu,\nu)}{2}\right) < 2 m(\alpha).
\label{condition_NK}
\end{equation}
Then $E^\nu(N)<E^\nu(N-K)+E^0(K)$ for any integer $K$ satisfying $K>N$ or $K<0$. Therefore, in this case the HVZ-type condition $(H_1')$ in Theorem \ref{HVZ} can be replaced by the more usual one
\begin{description}
\item[$(H_1'')$]  $E^\nu(N)<\min\{E^\nu(N-K)+E^0(K),\ K=1,...,N\}$.
\end{description}
\end{corollary}

The proof of Corollary \ref{simplification_HVZ} is left to the reader. When $m(\alpha)=g_0(0)$ (which is true when $\alpha\ll1$, see Lemma \ref{threshold}), \eqref{condition_NK} can be replaced by the stronger condition $\alpha\left((N+2)\pi/2+D(\nu,\nu)\right)< 4$.

\subsection{Approximation by finite-rank operators in $\gS_1^{\cP^0_-}$}
\begin{proposition}[Approximation by finite-rank operators]\label{finiterank_dense}
The set consisting of the operators $Q$ which satisfy
\begin{enumerate}
\item $Q\in\cQ_\Lambda(q)$;
\item $Q=P-\cP^0_-+\gamma$ where $P$ is an orthogonal projector and $\gamma$ is a finite rank operator such that $0\leq\gamma<1$, $P\gamma=\gamma P=0$;
\item $Q$ has a finite rank;
\end{enumerate}
is a dense subset of $\cQ_\Lambda(q)$ for the strong topology of $\gS_1^{\cP^0_-}$.
\end{proposition}

\begin{proof}
The proof relies on a useful parametrization of the variational set $\cQ_\Lambda$, presented and proved in Appendix B, Theorem \ref{thm_form_v_set}. This result itself is a generalization of a reduction in the case where $Q\in\cQ_\Lambda$ is a difference of two orthogonal projectors, see Theorem \ref{reduction}.
By Theorem \ref{thm_form_v_set}, any $Q\in\cQ_\Lambda(N)$ can be written 
$Q=U_D(\cP^0_-+\gamma)U_{-D}-\cP^0_-$
where $\gamma\in\gS_1(\gH_\Lambda)$, $[\gamma,\cP^0_-]=0$, $D\in\gS_2(\gH_\Lambda)$ and $U_D=\exp(D-D^*)$. Moreover $\gamma=\gamma^+-\gamma^-$ where $0\leq\gamma^+\leq \cP^0_+$ and $0\leq\gamma^-\leq \cP^0_-$. Clearly we can find sequences $\{D_n\}$, $\{\gamma^\pm_n\}$ of finite rank operators such that $D_n\to D$ in $\gS_2(\gH_\Lambda)$, $\gamma^\pm_n\to\gamma^\pm$ in $\gS_1(\cP^0_\pm\gH_\Lambda)$ as $n\to\ii$, and additionally $\tr(\gamma^+_n-\gamma^-_n)=\tr(\gamma)$ for any $n$. Then $U_{D_n}\cP^0_-U_{-D_n}- U_{D}\cP^0_-U_{-D}\to0$ in $\gS_2(\gH_\Lambda)$. We know from \cite[Lemma 2]{HLS1} that $$\tr_{\cP^0_-}(U_{D_n}\cP^0_-U_{-D_n}-\cP^0_-)=\tr(U_{D_n}\cP^0_-U_{-D_n}-\cP^0_-)^3$$
 is an integer. Thus $\tr_{\cP^0_-}(U_{D_n}\cP^0_-U_{-D_n}-\cP^0_-)$ is constant for $n$ large enough and
$$Q_n=U_{D_n}(\cP^0_-+\gamma^+_n-\gamma^-_n)U_{-D_n}-\cP ^0_-$$
is a sequence of finite rank operators which converges to $Q$ in $\cQ_\Lambda(q)$.
Assume now that 
$$\gamma^+_n=\sum_{i=1}^{m^+_n}\lambda_i^+|\phi_i^+\rangle\langle\phi_i^+|\quad \text{and }\quad \gamma^-_n=\sum_{i=1}^{m^-_n}\lambda^-_i|\phi^-_i\rangle\langle\phi^-_i|$$
with $0<\lambda_i^{\pm}\leq1$ and $\phi_i^\pm\in\cP^0_\pm\gH_\Lambda$. We then introduce
$$P_n':=\cP^0_--\sum_{i=1}^{m^-_n}|\phi^-_i\rangle\langle\phi^-_i|+\sum_{i\ |\ \lambda^+_i=1}|\phi^+_i\rangle\langle\phi^+_i|\quad\text{and}\quad\gamma'_n:=\cP^0_-+\gamma_n^+-\gamma_n^--P_n'.$$ 
Then $Q_n=U_{D_n}(P'_n+\gamma'_n)U_{-D_n}-\cP^0_-=\tilde P_n+\tilde\gamma_n-\cP^0_-$ satisfies the assumptions of the Proposition.
\qed \qed \end{proof}

\begin{corollary}
There exists a minimizing sequence $(Q_n)_{n\geq1}$ of $E^\nu(q)$, satisfying the three conditions of Proposition \ref{finiterank_dense}.
\end{corollary}

As usual in Hartree-Fock type theories \cite{Lieb,Bach,BLS}, we now prove that minimizing $\E^\nu$ in the convex set of states in $\cQ_\Lambda$ having charge $-eq$ is equivalent to minimizing on extremal points only.
\begin{proposition}[Lieb's variational principle]\label{lieb}
Let be $0\leq\alpha<4/\pi$, $\Lambda>0$, $\nu\in\cC$ and $q\in\R$. One has 
\begin{multline}
E^\nu(q) = \inf\big\{ \E^\nu(Q)\ |\ Q=P-\cP^0_-+(q-[q])|\psi\rangle\langle\psi|\in\cQ_\Lambda,\\
    P^2=P=P^*,\ \str(P-\cP^0_-)=[q],\ \psi\in\gH_\Lambda,\ P\psi=0\big\}.\label{egalite_lieb}
\end{multline}
\end{proposition}

\begin{proof}
 We use a well-known method for the study of Hartree-Fock type theories \cite{Lieb,Bach,BLS}. The inequality $\leq$ in \eqref{egalite_lieb} being trivial, it suffices to prove the converse. To this end, we consider by Proposition \ref{finiterank_dense} a state $Q=P-\cP^0_-+\gamma\in\cQ_\Lambda(q)$ such that  $P^2=P=P^*$, $\gamma$ is a finite-rank operator with $0\leq\gamma<1$, $P\gamma=\gamma P=0$, $\tr_{\cP^0_-}Q=q$, and
$\E^\nu(Q)\leq E^\nu(q)+\epsilon$.
If $\gamma$ has rank greater than 1, then one can find two orthogonal eigenfunctions $\chi_1$ and $\chi_2$ of $\gamma$ corresponding to eigenvalues $\gamma_1$ and $\gamma_2$ in $(0;1)$.
Next one can compute the energy of $Q+t(|\chi_2\rangle\langle\chi_2| - |\chi_1\rangle\langle\chi_1|)$
which belongs to $\cQ_\Lambda(q)$ as soon as $\gamma_2+t\in[0;1]$ and $\gamma_1-t\in[0;1]$. One sees that, depending on the sign of $\pscal{\cD_Q\chi_2,\chi_2}-\pscal{\cD_Q\chi_1,\chi_1}$, we can decrease the energy by increasing or decreasing $t$ until $\gamma_2+t\in\{0;1\}$ or $\gamma_1-t\in\{0;1\}$. Doing so, we obtain a new state 
$\bar Q=\bar P-\cP^0_-+\bar\gamma$
where $\bar P$ is an orthogonal projector and $\bar \gamma$ is such that $0\leq\bar\gamma<1$, $\bar P\bar\gamma=\bar\gamma\bar P=0$ and ${\rm rank}(\bar\gamma)\leq {\rm rank}(\gamma)-1$. Iterating this process, we can eliminate in finitely many steps all the eigenvalues in $(0;1)$ of the finite-rank operator $\gamma$, except possibly one, and decrease the energy at each step. We end up with a state $Q'=P'-\cP^0_-+\lambda|\psi\rangle\langle\psi|$ where $\lambda\in[0;1)$, $P'\psi=0$ and $\E^\nu(Q')\leq  E^\nu(q)+\epsilon$.
It then suffices to use \cite[Lemma 2]{HLS1} which tells us that $\str(P'-\cP^0_-)$ is always an integer, to conclude that $\lambda=q-[q]$.
\qed \end{proof}

\subsection{HVZ-type inequalities}
We use the density of finite-rank operators to prove HVZ-type inequalities.
\begin{proposition}[HVZ-type inequalities]\label{HVZ_large}
Let be $0\leq\alpha<4/\pi$, $\Lambda>0$, $\nu\in\cC$ and $q\in\R$. Then one has
\begin{equation}
\label{large}
E^\nu(q) \leq \min\left\{ E^\nu(q-k)+E^0(k),\ k\in\R\right\}.
\end{equation}
If moreover $E^\nu(q)=E^\nu({q-k})+E^0(k)$ for some $k\neq0$, then there exists a minimizing sequence $(Q_n)_{n\geq1}$ of $E^\nu(q)$ which satisfies $Q_n\wto Q$ in $\cQ_\Lambda$ with $\str(Q)=q-k$, and which is therefore not precompact in $\cQ_\Lambda(q)$.
\end{proposition}

\begin{proof}
Notice that $E^0(0)=0$ by \cite[Theorem 1]{HLS1}, so there is nothing to prove when $k=0$. Let us fix some $k\neq0$ and consider two states $Q,\ Q'\in\cQ_\Lambda$ such that
$\E^0(Q)\leq E^0(k)+\epsilon$ and $\E^\nu(Q')\leq E^\nu(q-k)+\epsilon$,  $\epsilon>0$.
Using Proposition \ref{finiterank_dense}, Proposition \ref{lieb} and Theorem \ref{reduction}, we can choose $Q$ of the following form:
\begin{eqnarray}
Q & = & \lambda|\psi\rangle\langle\psi|+\sum_{n=1}^{N_1} |f_n\rangle\langle f_n|-\sum_{m=1}^{N_2} |g_m\rangle\langle g_m|  +\sum_{i=1}^K\frac{\lambda_i^2}{1+\lambda_i^2}\big(|v_i\rangle\langle v_i|-|u_i\rangle\langle u_i|\big)\nonumber\\
 &  & \qquad +\sum_{i=1}^K\frac{\lambda_i}{1+\lambda_i^2}\big(|u_i\rangle\langle v_i|+|v_i\rangle\langle u_i|\big)\label{Q_app_1}.
\end{eqnarray}
where $\lambda=k-[k]\in[0;1)$, , $N_1-N_2=[k]$.
In \eqref{Q_app_1} $(f_i)_{i=1}^{N_1}\cup(v_i)_ {i=1}^K$ forms an orthonormal set of $\cP^0_+\gH_\Lambda$ and $(g_i)_{i=1}^{N_2}\cup(u_i)_{i=1}^{K}$ is an orthonormal set of $\cP^0_-\gH_\Lambda$.
We may choose the operator $Q'$ of a similar form (with functions $\psi'$, $f'_n$, $g_m'$, $u'_i$, $v'_i$ and real numbers $\lambda'$, $\lambda'_i$).

The free minimization problem $E^0(k)$ being translation-invariant, we have $\E^0(\tau_tQ'\tau_t^*)=\E^0(Q')$ for any $t\in\R$, where $\tau_t$ is the unitary operator $\tau_tf(x)=f(x-te)$, $e$ being a fixed unitary vector in $\R^3$. Notice $\tau_tQ'\tau_t^*\in\cQ_\Lambda$, since both $\cP^0_-$ and $\cP^0_+$ are translation-invariant.
Since $\lim_{t\to\ii}\pscal{\tau_tf,g}=0$ for any $f,g\in\gH_\Lambda$, we can find for $t\gg1$ by the Gram-Schmidt orthonormalization procedure two orthonormal systems 
$(f^t_n)_{n=1}^{N_1}\cup(v^t_i)_ {i=1}^{K}$ in the orthogonal of ${\rm span}\{f'_n,\ v'_i\}$ inside $\cP^0_+\gH_\Lambda$, and 
$(g^t_i)_{i=1}^{N_2}\cup(u^t_i)_ {i=1}^{K}$ in the orthogonal of ${\rm span}\{g'_m,\ v'_i\}$ inside $\cP^0_-\gH_\Lambda$
which are such that $\lim_{t\to\ii}\norm{f^t_n- \tau_tf_n}_{\gH_\Lambda}=0$ and a similar properties for all the other functions. Substituting in \eqref{Q_app_1}, this defines us an operator $Q^t\in\cQ_\Lambda$ such that $\lim_{t\to\ii}\norm{Q^t-\tau_tQ\tau_t^* }_{\gS_1(\gH_\Lambda)}=0$. Moreover, we have by construction $Q^t+Q'\in\cQ_\Lambda$ and $\tr_{\cP^0_-}(Q^t+Q')=q$. It can be seen that
\begin{multline}
 E^\nu(q)\leq \E^\nu(Q^t+Q')= \E^\nu(Q')+\E^0(Q)+o_{t\to\ii}(1)\\ \leq E^\nu(q-k)+E^0(k)+2\epsilon+o_{t\to\ii}(1),
\end{multline}
which, passing to the limit as $t\to\ii$ and $\epsilon\to0^+$, ends the proof of this first part.
If $E^q(\phi)=E^{q-k}(\phi)+E^k(0)$ for some $k\neq0$, then one constructs a non-compact minimizing sequence by the same argument.
\qed \end{proof}

\begin{corollary}\label{continuity_E} Let be $0\leq\alpha< 4/\pi$, $\nu\in\cC$ and $\Lambda>0$. Then the map $q\mapsto E^\nu(q)$ is uniformly Lipschitz on $\R$:
$$|E^\nu(q)-E^\nu(q')|\leq g_0(0)\,|q-q'|.$$
\end{corollary}
\begin{proof}
This is an obvious consequence of \eqref{large} and \eqref{encadrement_min}.
\qed \end{proof}

%%%%%%%%%%%%%%%%%%%%%%%%%%%%%%%%%%%%
\section{Proof of Theorem \ref{HVZ}}
We prove that $(H_1)$ implies $(H_2)$, as the converse was shown in Proposition \ref{HVZ_large}.

\medskip

\noindent{\bf Step 1 : }{\it Reduction to the HVZ condition for integers when $q=N\in\Z$}. 
\begin{lemma}\label{HVZ_integer} Let be $0\leq\alpha< 4/\pi$, $\nu\in\cC$ and $N\in\Z$. Then for any $K\in\Z$, the function  $k\mapsto E^\nu(N-k)+E^0(k)$ is concave on $[K,K+1]$.
Therefore, if $(H_1')$ holds, then so does $(H_1)$.
\end{lemma}

\begin{proof}
We prove that the function  $k\mapsto E^\nu(k)$ is concave on $[K,K+1]$, the rest being obvious. This means that we prove for any $k_1,k_2,\mu\in[0,1]$
\begin{equation}
E^\nu(K+\mu k_1+(1-\mu)k_2)\geq \mu E^\nu(K+ k_1)+(1-\mu)E^\nu(K+ k_2).
\end{equation}
To this end, let us consider like in the proof of Proposition \ref{HVZ_large}, one state $Q$ which satisfies $\E^\nu(Q)  \leq  E^\nu(K+\mu k_1+(1-\mu)k_2)+\epsilon$ for some fixed $\epsilon>0$.
By Proposition \ref{lieb}, we may take $Q=(\mu k_1+(1-\mu)k_2)|\psi\rangle\langle\psi|+P-\cP^0_-$
where $P$ is an orthogonal projector satisfying $P\psi=0$ and $\tr_{\cP^0_-}(P-\cP^0_-)=K$.
We have
\begin{align*}
\E^\nu(Q) & = (\mu k_1+(1-\mu)k_2)\pscal{\cD_{[P-\cP^0_-]}\psi,\psi}+\E^\nu(P-\cP^0_-)\\
 & = \mu\E^\nu(P-\cP^0_-+k_1|\psi\rangle\langle\psi|)+(1-\mu)\E^\nu(P-\cP^0_-+k_2|\psi\rangle\langle\psi|)\\
 & \geq \mu E^\nu(K+k_1) +(1-\mu)E^\nu(K+k_2)
\end{align*}
where we have used that an electron never sees its own field.
This ends the proof, taking $\epsilon\to0$.
\qed \end{proof}

\noindent{\bf Step 2: } {\it A necessary and sufficient condition for compactness}. We now assume that $q\in\R$ and prove that conservation of the charge implies the compactness of minimizing sequences.

\begin{lemma}\label{trace_compact} Let be $0\leq\alpha< 4/\pi$, $\Lambda>0$, $\nu\in\cC$ and $q\in\R$. Assume that $(Q_n)_{n\geq1}$ is a minimizing sequence of $E^\nu(q)$ such that $Q_n\wto Q$ for the weak topology of $\cQ_\Lambda$. Then $Q_n\to Q$ for the strong topology of $\cQ_\Lambda$ if and only if $\str(Q)=q$.
\end{lemma}

\begin{proof}
By assumption, we have $\lim_{n\to\ii}\E^\nu(Q_n)=E^\nu(q)$. If $\str(Q)=q$, then $Q$ is a minimizer for $E^\nu(q)$ since $\E^\nu$ is weakly lower semi-continuous. Also we have \cite{HLS2}
$\lim_{n\to\ii}\norm{\rho_{Q_n}-\rho_Q}_{\cC}=0\ \text{ and }\ \lim_{n\to\ii} F(Q_n)=F(Q)$
where
$$F(Q):= \str(\cD^0Q)-\frac{\alpha}{2}\iint_{\R^6}\frac{|Q(x,y)|^2}{|x-y|}dx\,dy.$$

Let us first prove that $Q_n\to Q$ in $\gS_2(\gH_\Lambda)$, or equivalently $\tr(Q_n^2)\to\tr(Q^2)$.  
To this end, we argue like in the proof of \cite[Theorem 1]{HLS2} and consider two smooth functions $\chi,\xi\in C^\ii([0;\ii),[0;1])$ such that $\chi^2+\xi^2=1$ and
$$\chi(x)=\left\{\begin{array}{cll}
= & 1 & \text{when } x\in[0;1]\\
\in & [0;1] & \text{when } x\in[1;2]\\
= & 0 & \text{when } x\geq2.
\end{array}\right.$$
We then define $\chi_R(x):=\chi(|x|/R)$ and $\xi_R(x):=\xi(|x|/R)$ for $x\in\R^3$.
As shown in \cite[page 4494]{HLS2}, one has
\begin{multline}
F(Q_n)=F(Q)+\tr(|\cD^0|\xi_R(Q_n^{++}-Q_n^{--})\xi_R)\\-\frac\alpha2\iint_{\R^6}\frac{\xi_R(x)^2|Q_n(x,y)|^2}{|x-y|}dx\,dy+\epsilon_n^R
\label{Decomp_energy}
\end{multline}
where $\epsilon_n^R$ satisfies
$\lim_{R\to\ii}\limsup_{n\to\ii}|\epsilon_n^R|=0\label{limit_epsilon}$
(see the details in \cite{HLS2}). Localizing the inequality $Q_n^2\leq Q_n^{++}-Q_n^{--}$ and applying Kato's inequality $|x|^{-1}\leq \pi/2|\nabla|$ like in \cite{HLS2} together with $|\nabla|\leq |D^0|\leq |\cD^0|$ as shown in \cite{HLSo}, we get
\begin{multline}
\tr(|\cD^0|\xi_R(Q_n^{++}-Q_n^{--})\xi_R)-\frac\alpha2\iint_{\R^6}\frac{\xi_R(x)^2|Q_n(x,y)|^2}{|x-y|}dx\,dy\\
\geq (1-\alpha\pi/4)\tr(|\cD^0|\xi_RQ_n^2\xi_R)\geq (1-\alpha\pi/4)\tr(\xi_RQ_n^2\xi_R).
\end{multline}
By \eqref{Decomp_energy}, \eqref{limit_epsilon} and $F(Q_n)\to F(Q)$, this proves that 
$$\lim_{R\to\ii}\limsup_{n\to\ii}\tr(\xi_RQ_n^2\xi_R)=0,$$
when $0\leq \alpha<4/\pi$. On the other hand
$$\tr(Q_n^2)=\tr(\chi_RQ_n^2\chi_R)+\tr(\xi_RQ_n^2\xi_R)$$
and 
$$\lim_{n\to\ii}\tr(\chi_RQ_n^2\chi_R)=\tr(\chi_RQ^2\chi_R)$$
for any fixed $R$, by the local compactness of $Q_n$. This implies
$$\lim_{n\to\ii}\tr(Q_n^2)=\tr(Q^2)$$
or equivalently $Q_n\to Q$ in $\gS_2(\gH_\Lambda)$.

It remains to prove that $Q_n^{++}$ and $Q_n^{--}$ converge strongly in the trace-class. We use the continuity of the exchange term for the Hilbert-Schmidt norm
$$\iint_{\R^6}\frac{|R(x,y)|^2}{|x-y|}dx\,dy\leq \frac\pi2\tr(|\nabla|R^2)\leq\frac{\Lambda\pi}2\norm{R}_{\gS_2(\gH_\Lambda)}^2,$$
due to Kato's inequality and the cut-off in Fourier space, to infer 
$$\lim_{n\to\ii}\iint_{\R^6}\frac{|Q_n(x,y)|^2}{|x-y|}dx\,dy=\iint_{\R^6}\frac{|Q(x,y)|^2}{|x-y|}dx\,dy.$$
This proves 
$$\lim_{n\to\ii}\tr(|\cD^0|(Q_n^{++}-Q_n^{--})=\tr(|\cD^0|(Q^{++}-Q^{--})$$
and therefore $Q_n^{++}\to Q^{++}$ and $Q_n^{--}\to Q^{--}$ in $\gS_1(\gH_\Lambda)$. 
\qed \end{proof}

We finish the proof of Theorem \ref{HVZ} by a contradiction argument: we assume that $(H_1)$ holds 
 and that there exists a minimizing sequence $(Q_n)_{n\geq1}$ of $E^\nu(q)$ which is not precompact for the strong topology of $\gS_1^{\cP^0_-}(\gH_\Lambda)$. Since $\E^\nu(Q_n)$ converges to $E^\nu(q)$, $(Q_n)$ is a bounded sequence in $\gS_1^{\cP^0_-}(\gH_\Lambda)$ and we may therefore assume, up to a subsequence, that $Q_n\wto Q$ for the weak topology of $\gS_1^{\cP^0_-}(\gH_\Lambda)$, and that $Q_n\nrightarrow Q$. By Lemma \ref{trace_compact}, this is equivalent to assuming that $\str(Q)\neq q$. We now write $\str(Q)=q-k$ with $k\in\R\setminus\{0\}$ and prove that this would imply $E^\nu(q)\geq E^\nu(q-k)+E^0(k)$,
which contradicts $(H_1)$.

As usual in HVZ or concentration-compactness type arguments, the rest of the proof now proceeds by decomposing the sequence $(Q_n)$ into a compact part converging strongly to $Q$ and a non-compact part escaping to infinity with the charge $k$. The localization of $Q$ is complicated by the constraint appearing in the definition of the variational space $\cQ_\Lambda(q)$, and the fact that our states are not bounded in the trace-class.

\bigskip

\noindent{\bf Step 3: }{\it The localization operators}. We introduce
$$X_R:= \cP^0_-\chi_R \cP^0_- +  \cP^0_+\chi_R \cP^0_+$$
and $Y_R$ which is the unique non-negative operator satisfying
$X_ R^2+ Y_R^2=1_{\gH_\Lambda}$.
Recall that the function $\chi_R$ has been defined in the previous step.
A crucial fact will be that both $X_R$ and $Y_R$ commute with $\cP^0_-$. The next three lemmas summarize useful properties of $X_R$ and $Y_R$.
\begin{lemma}[Continuity of the localization maps]\label{continuity_XR}
There exists a constant  $C$ independent on $R$ and $\Lambda$ such that, for any $Q\in\gS_1^{\cP^0_-}(\gH_\Lambda)$,
\begin{equation}
\label{estim_norm_XR}
\norm{X_RQX_R}_{1;\cP^0_-}+\norm{Y_RQY_R}_{1;\cP^0_-} \leq C \norm{Q}_{1;\cP^0_-}.
\end{equation}
If moreover $Q\in\cQ_\Lambda$, then $X_RQX_R$ and $Y_RQY_R$ also belong to $\cQ_\Lambda$. 
\end{lemma}

\begin{proof}
We notice that $(X_R)$ and $(Y_R)$ are uniformly bounded in $\gS_\ii(\gH_\Lambda)$. 
%%Début corrections
Indeed, using $0\leq\chi_R\leq 1$, one sees that $\norm{X_R}\leq 1$ and $\norm{Y_R}\leq 1$.
Therefore 
$$\norm{X_RQX_R}_{\gS_2(\gH_\Lambda)}\leq \norm{X_R}^2\norm{Q}_{\gS_2(\gH_\Lambda)}\leq \norm{Q}_{\gS_2(\gH_\Lambda)},$$
%%FIn corrections
and, using that $X_R$ commutes with $\cP^0_-$,
$$\norm{(X_RQX_R)^{--}}_{\gS_1(\gH_\Lambda)}  =  \norm{X_RQ^{--}X_R}_{\gS_1(\gH_\Lambda)}
  \leq \norm{Q^{--}}_{\gS_1(\gH_\Lambda)}$$
by the same argument as above. Using the same idea for $Y_RQY_R$, one obtains \eqref{estim_norm_XR}.
Let us now prove that if $Q$ satisfies the constraint $-\cP^0_-\leq Q\leq \cP^0_+$, then so does $X_RQX_R$, the argument being the same for  $Y_RQY_R$. We have
$ X_RQX_R\leq X_R\cP^0_+X_R=(\cP^0_+\chi_R\cP^0_+)^2\leq \cP^0_+\chi_R^2\cP^0_+\leq\cP^0_+.$
A similar argument for $X_RQX_R\geq-\cP^0_-$ allows to end the proof of Lemma \ref{continuity_XR}.
\qed \end{proof}

%%Début Corrections
\begin{lemma}[Limits as $R\to\ii$]\label{lim_XR_lemma}
One has
\begin{equation}
\label{lim_XYR}
\lim_{R\to\ii}\norm{X_R-\chi_R}=0, \qquad \lim_{R\to\ii}\norm{Y_R-\xi_R}=0.
\end{equation}
Moreover, if $Q\in\gS_1^{\cP^0_-}(\gH_\Lambda)$, then 
\begin{equation}\label{lim_XR_pp}
\lim_{R\to\ii}\norm{X_RQX_R-Q}_{1;\cP^0_-}=\lim_{R\to\ii}\norm{Y_RQY_R}_{1;\cP^0_-}=0.
\end{equation}
\end{lemma}

\begin{proof}
To prove \eqref{lim_XYR} for $X_R$, we compute
$X_R-\chi_R=[\cP^0_-,\chi_R]\cP^0_-+[\cP^0_+,\chi_R]\cP^0_+$
which tends to 0 as $R\to\ii$ by \cite[Lemma 1]{HLS2} and the proof of Theorem 3 in \cite{HLSo}. This clearly implies that $\lim_{R\to\ii}\norm{X_R^2-\chi_R^2}=0$, since $\norm{X_R}\leq1$ and $\norm{\chi_R}\leq1$.
We now use the fact that the square root is operator monotone to deduce \cite[Thm X.1.1]{Bhatia}
$$\norm{Y_R-\xi_R}\leq \norm{Y_R^2-\xi_R^2}^{1/2}=\norm{X_R^2-\chi_R^2}^{1/2}$$
which proves \eqref{lim_XYR} for $Y_R$.
By the uniform boundedness of $(X_R)_R$ in $\gS_\ii(\gH_\Lambda)$, we can prove \eqref{lim_XR_pp} for $Q$ in a dense subset of $\gS_1^{\cP^0_-}(\gH_\Lambda)$, like finite-rank operators. By linearity, it suffices to prove \eqref{lim_XR_pp} for a state of the form $Q=|\psi\rangle\langle\psi|$. Using now \eqref{lim_XYR}, it remains to prove that $\chi_R|\psi\rangle\langle\psi|\chi_R-|\psi\rangle\langle\psi|$ converges to 0 in $\gS_1(\gH_\Lambda)$, which is a triviality since $\chi_R\psi\to\psi$ as $R\to\ii$ in $\gH_\Lambda$, by Lebesgue's dominated convergence theorem. 
We argue similarly for $Y_RQY_R$. 
\qed \end{proof}
%%Fin Corrections

\begin{lemma}[Compactness for a fixed $R$] \label{prop_LR}
For any fixed $R$, $X_R$  and $1-Y_R$ belong to $\gS_1(\gH_\Lambda)$ and are therefore compact. 
The map $Q\mapsto X_RQX_R$ is also compact: if $Q_n\wto Q$ for the weak topology of $\gS_1^{\cP^0_-}(\gH_\Lambda)$, then $X_RQ_nX_R\to X_RQX_R$ strongly in $\gS_1^{\cP^0_-}(\gH_\Lambda)$. The same holds if $X_R$ is replaced by $1-Y_R$.
\end{lemma}

\begin{proof}
We use the Kato-Seiler-Simon inequality (see \cite{SeSi} and \cite[Theorem 4.1]{Simon}) 
$$\norm{f(-i\nabla)g(x)}_{\gS_2}\leq C\norm{f}_{L^2}\norm{g}_{L^2}$$
to obtain
\begin{equation*}
\norm{X_R}_{\gS_1(\gH_\Lambda)} \leq \norm{\cP^0_-\sqrt{\chi_R}}_{\gS_2(\gH_\Lambda)}^2+\norm{\cP^0_+\sqrt{\chi_R}}_{\gS_2(\gH_\Lambda)}^2
 \leq 2C\,|B(0,\Lambda)|\, R\int_{\R^3}\chi
\end{equation*}
which is finite for each fixed $R$.
Notice that 
$0\leq 1-Y_R\leq 1-Y_R^2=X_ R^2\leq X_R$
and therefore $1-Y_R$ is also in $\gS_1(\gH_\Lambda)$.
Let us now prove that when $Q_n\wto Q$ for the weak topology of $\gS_1^{\cP^0_-}(\gH_\Lambda)$, then $X_RQ_nX_R\to X_RQX_R$ strongly in $\gS_1^{\cP^0_-}(\gH_\Lambda)$. Since $X_R$ is in $\gS_1(\gH_\Lambda)$ and by density of finite-rank operators in the trace-class, it suffices to prove that $|\phi\rangle\langle\phi|\ Q_n\,|\psi\rangle\langle\psi|\to |\phi\rangle\langle\phi|\,Q\,|\psi\rangle\langle\psi|$, for some $\phi$ and $\psi$ in $\gH_\Lambda$. This is a consequence of 
$|\phi\rangle\langle\phi|\ Q_n\,|\psi\rangle\langle\psi|=\tr(Q_n|\psi\rangle\langle\phi|)|\phi\rangle\langle\psi|$
and 
$\lim_{n\to\ii}\tr(Q_n|\psi\rangle\langle\phi|)=\tr(Q|\psi\rangle\langle\phi|)$
by the weak convergence of $Q_n$ in $\gS_2(\gH_\Lambda)$. The proof is the same for $1-Y_R$.
\qed \end{proof}

\smallskip

\noindent{\bf Step 4: }{\it Conclusion.} Let us recall that we argue by contradiction: we assume that $(H_1)$ holds and that $(Q_n)\subset \cQ_\Lambda$ is a minimizing sequence for $E^\nu(q)$ satisfying $Q_n\wto Q$ with $\str(Q)=q-k$, $k\neq0$. 
We fix some $R$ and compute
\begin{multline*}
\str(\cD^0Q_n) =\str(\cD^0(X_RQ_nX_R))+\str(\cD^0(Y_RQ_nY_R))\\
+\tr([X_R,|\cD^0|](Q_n^{++}-Q_n^{--})X_R)+\tr([Y_R,|\cD^0|](Q_n^{++}-Q_n^{--})Y_R),
\end{multline*}
where we have  used that $X_R$ and $Y_R$ commute with $\cP^0_-$.
Using now
$$|\tr([X_R,|\cD^0|](Q_n^{++}-Q_n^{--})X_R)|\leq C \norm{[X_R,|\cD^0|]}_{\cB(\gH_\Lambda)}$$
due to the fact that $(X_R)$ is uniformly bounded in $\cB(\gH_\Lambda)$ and $(Q_n^{++}-Q_n^{--})$ is uniformly bounded in $\gS_1(\gH_\Lambda)$, we obtain for some uniform constant $C$
\begin{align}
\str(\cD^0Q_n) &\geq \str(\cD^0Q)+\str(\cD^0(Y_RQ_nY_R))\nonumber\\
&\quad+\str(\cD^0(X_R(Q_n-Q)X_R))+\str(\cD^0(X_RQX_R-Q))\nonumber\\
&\quad-C\big(\norm{[X_R,|\cD^0|]}+\norm{[Y_R,|\cD^0|]}\big).\label{estim_str_below}
\end{align}

To treat the direct and exchange terms, we shall need the following 
\begin{lemma}\label{lemme_YR} Assume that $(R_n)$ is a sequence in $\gS_1^{\cP^0_-}(\gH_\Lambda)$ with $R_n\wto0$ as $n\to\ii$, for the weak topology of $\gS_1^{\cP^0_-}(\gH_\Lambda)$. Then, for any fixed $R$,
\begin{equation}
\label{limit_1}
\lim_{n\to\ii}\norm{\rho_{R_n}-\rho_{Y_RR_nY_R}}_{\cC}=0,
\end{equation}
\begin{equation}
\label{limit_2}
\lim_{n\to\ii}\iint_{\R^6}\frac{|(R_n-Y_RR_nY_R)(x,y)|^2}{|x-y|}dx\,dy=0.
\end{equation}
\end{lemma}

\begin{proof}
Recall that the exchange term is continuous for the $\gS_1^{\cP^0_-}(\gH_\Lambda)$ topology.
We write
$$R_n-Y_RR_nY_R=-(1-Y_R)R_n(1-Y_R)+R_n(1-Y_R)+(1-Y_R)R_n.$$
By Lemma \ref{prop_LR}, $(1-Y_R)R_n(1-Y_R)$ converges strongly to 0 in the $\gS_1^{\cP^0_-}(\gH_\Lambda)$ norm. It therefore suffices to prove \eqref{limit_1} and \eqref{limit_2} with $R_n-Y_RR_nY_R$ replaced by  
$R_n(1-Y_R)$ and $(1-Y_R)R_n$. The operator $(1-Y_R)$ being trace-class by Lemma \ref{prop_LR}, we can prove 
\eqref{limit_1} and \eqref{limit_2} with $R_n-Y_RR_nY_R$ replaced by $S_n:=R_n\,|\phi\rangle\langle\phi|\in\gS_1(\gH_\Lambda)$ for some $\phi\in\gH_\Lambda\cap L^1(\R^3)\subset L^2(\R^3,\C^4)$. We have
$$\rho_{S_n}(x)=\int_{\R^3}\tr_{\C^4}\left(R_n(x,y)\phi(y)\phi(x)^*\right)dy$$
and therefore
$$\norm{\rho_{S_n}}_{L^1} \leq \iint_{\R^6}|R_n(x,y)|\,|\phi(x)|\,|\phi(y)|dx\,dy.$$
Thanks to the cut-off in Fourier space, we can assume, up to a subsequence, that $R_n(x,y)$ converges uniformly to 0 on compact subsets of $\R^6$. Since we have assumed that $\phi\in L^1(\R^3)$, we conclude by Lebesgue's dominated convergence theorem that $\lim_{n\to\ii}\norm{\rho_{S_n}}_{L^1}=0$, which itself implies $\lim_{n\to\ii}\norm{\widehat{\rho_{S_n}}}_{L^\ii}=0$ and $\lim_{n\to\ii}\norm{\rho_{S_n}}_{\cC}=0$ thanks to the cut-off in Fourier space.

We use the same argument for the exchange term:
\begin{equation*}
\iint_{\R^6}\frac{|S_n(x,y)|^2}{|x-y|}dx\,dy \leq \iint_{\R^6}|R_n(x,z)|^2|\phi(z)|^2\left(|\phi|^2\ast\frac{1}{|\cdot|}\right)(x)dx\,dz
\end{equation*}
which converges to 0 as $n\to\ii$ by the local compactness of $(R_n)$. This ends the proof of Lemma \ref{lemme_YR}
\qed \end{proof}

We are now able to finish the proof of Theorem \ref{HVZ}. We write
\begin{multline}
D(\rho_{Q_n},\rho_{Q_n})  =  D(\rho_{Q},\rho_{Q})+D(\rho_{Y_R(Q_n-Q)Y_R},\rho_{Y_R(Q_n-Q)Y_R})+\epsilon_1^R(n)\\
 \geq  D(\rho_{Q},\rho_{Q})+D(\rho_{Y_RQ_nY_R},\rho_{Y_RQ_nY_R}) +\epsilon_1^R(n)-C_1\norm{\rho_{Y_RQY_R}}^2_\cC,\label{decom_rho}
\end{multline}
\begin{align}
&\iint_{\R^6}\frac{|Q_n(x,y)|^2}{|x-y|}dx\,dy\nonumber\\
 &\quad = \iint_{\R^6}\frac{|Q(x,y)|^2}{|x-y|}dx\,dy+\iint_{\R^6}\frac{|(Y_R(Q_n-Q)Y_R)(x,y)|^2}{|x-y|}dx\,dy+\epsilon_3^R(n)\nonumber\\
  &\quad \leq  \iint_{\R^6}\frac{|Q(x,y)|^2}{|x-y|}dx\,dy +\iint_{\R^6}\frac{|(Y_RQ_nY_R)(x,y)|^2}{|x-y|}dx\,dy\nonumber\\
  &\qquad\qquad \qquad +\epsilon_2^R(n) +C_2\iint_{\R^6}\frac{|Y_RQY_R(x,y)|^2}{|x-y|}dx\,dy,\label{decom_Q}
\end{align}
and
\begin{equation}
\label{decom_rho2}
D(\rho_{Q_n},\nu)=D(\rho_{Q},\nu)+\epsilon_3(n).
\end{equation}
In \eqref{decom_rho}, \eqref{decom_rho2} and \eqref{decom_Q}, $C_1$ and $C_2$ are uniform constants (we have used that $(Q_n)$ is bounded in $\gS_1^{\cP^0_-}(\gH_\Lambda)$), and $\epsilon_3(n):=D(\rho_{Q_n-Q},\nu),$
$$\epsilon_1^R(n):=\norm{\rho_{Q_n-Q}}_\cC^2-\norm{\rho_{Y_R(Q_n-Q)Y_R}}^2_{\cC}+2D(\rho_{Q_n-Q},\rho_Q),$$
\begin{multline}
\epsilon_2^R(n):=\iint_{\R^6}\frac{|(Q_n-Q)(x,y)|^2}{|x-y|}dx\,dy-\iint_{\R^6}\!\!\!\frac{|Y_R(Q_n-Q)Y_R(x,y)|^2}{|x-y|}dx\,dy\\
+2\Re\iint_{\R^6}\frac{\tr_{\C^4}[(Q_n-Q)(x,y)\overline{Q(x,y)}]}{|x-y|}dx\,dy
\end{multline}
satisfy, for each fixed $R$,
$\lim_{n\to\ii}\epsilon_1^R(n)=\lim_{n\to\ii}\epsilon_2^R(n)=\lim_{n\to\ii}\epsilon_3(n)=0,$
by Lemma \ref{lemme_YR} applied to $R_n=Q_n-Q$.
By \eqref{estim_str_below}, \eqref{decom_rho}, \eqref{decom_rho2} and \eqref{decom_Q}, we obtain
\begin{align}
\E^\nu(Q_n)&\geq \E^\nu(Q)+\E^0(Y_RQ_nY_R)+\epsilon^R_1(n)+\epsilon_2(n)-\epsilon^R_3(n)\nonumber\\
&\qquad+\str(\cD^0(X_R(Q_n-Q)X_R))+\str(\cD^0(X_RQX_R-Q))\nonumber\\
&\qquad\qquad-C\norm{[Y_R,|\cD^0|]} -C\norm{[X_R,|\cD^0|]}-C_1\norm{\rho_{Y_RQY_R}}^2_\cC\nonumber\\
&\qquad\qquad\qquad-C_2\iint_{\R^6}\frac{|Y_RQY_R(x,y)|^2}{|x-y|}dx\,dy.\label{estim_below_energy1}
\end{align}
Let us remark that 
$q=\str(Q_n)=\str(X_RQ_nX_R)+\str(Y_RQ_nY_R)$
where we have used $[X_R,\cP^0_-]=0$. In \eqref{estim_below_energy1}, we can estimate
\begin{eqnarray*}
\E^\nu(Q)+\E^0(Y_RQ_nY_R) & \geq & E^\nu(q-k)+E^0(\str(Y_RQ_nY_R))\\
 & = & E^\nu(q-k)+E^0(q-\str(X_RQ_nX_R)).
\end{eqnarray*}
Passing to the limit $n\to\ii$ in \eqref{estim_below_energy1} with $R$ fixed and using Lemma \ref{prop_LR} together with the continuity of $q\mapsto E^\nu(q)$ as stated in Corollary \ref{continuity_E}, we get
\begin{align}
E^\nu(q)&\geq E^\nu(q-k)+E^0(q-\str(X_RQX_R))+\str(\cD^0(X_RQX_R-Q))\nonumber\\
&\qquad-C\norm{[X_R,|\cD^0|]}-C\norm{[Y_R,|\cD^0|]}\nonumber\\
&\qquad\qquad-C_1\norm{\rho_{Y_RQY_R}}^2_\cC-C_2\iint_{\R^6}\frac{|Y_RQY_R(x,y)|^2}{|x-y|}dx\,dy.\label{estim_below_energy2}
\end{align}
Note that 
\begin{multline*}
\lim_{R\to\ii}\norm{[X_R,|\cD^0|]}=\lim_{R\to\ii}\norm{[\chi_R,|\cD^0|]}=\lim_{R\to\ii}\norm{[Y_R,|\cD^0|]}\\
=\lim_{R\to\ii}\norm{[\xi_R,|\cD^0|]}=0
\end{multline*}
by Lemma \ref{lim_XR_lemma} together with \cite[Lemma 1]{HLS2}, \cite[Proof of Thm 3]{HLSo} and the boundedness of $\cD^0$ on $\gH_\Lambda$.
Passing to the limit as $R\to\ii$ we eventually obtain from Lemma \ref{lim_XR_lemma}
$$E^\nu(q)\geq E^\nu(q-k)+E^0(q-\str(Q))=E^\nu(q-k)+E^0(k).$$
This contradicts $(H_1)$ and ends the proof of Theorem \ref{HVZ}.\qed
%%Fin corrections

%%%%%%%%%%%%%%%%
\section{Proof of Theorem \ref{exists_weak_limit}}
Note that when $\bar\nu\in\cC$, the essential spectrum of $D^0-\bar\nu\ast|\cdot|^{-1}$ is the same as the one of $D^0$.
 Assuming that $\ker(D^0-\bar\nu\ast|\cdot|^{-1})=\{0\}$, we denote by $(\lambda_i^+)_{i\geq1}$ the non-decreasing sequence of eigenvalues of $D^0-\bar\nu\ast|\cdot|^{-1}$ in $(0,1)$, counted with their multiplicity. In case there is a finite number $i_0$ of eigenvalues or no eigenvalue at all ($i_0=0$) in $(0,1)$, we then let $\lambda_{i_0+1}^+=1$ and $\lambda_{i}^+=1$ for $i\geq i_0+1$. We use the same type of notation $(\lambda_i^-)_{i\geq1}$ for the non-increasing sequence of eigenvalues in $(-1,0)$, with $\lambda_{i_0'+1}^-=-1$ in case there is a finite number of (possibly no) eigenvalues in $(-1,0)$.
 We notice that $\ker(D^0-\bar\nu\ast|\cdot|^{-1})=\{0\}$ implies that there exists some constant $\kappa>0$ such that
\begin{equation}
|D^0-\bar\nu\ast|\cdot|^{-1}|\geq \kappa.
\label{estim_below_linear}
\end{equation}
 
 \medskip
 
 \noindent{\bf Step 1:} {\it Study of the linear model.} We start by computing the value of the infimum in the right hand side of \eqref{limit_weak_coupling}, in terms of the eigenvalues $(\lambda_i^\pm)$ of $D^0-\bar\nu\ast|\cdot|^{-1}$, and the charge of the vacuum. We define
 \begin{equation}
I^{\bar\nu}(N):=\inf_{\substack{Q\in\gS^{P^0_-}_1(\gH_\Lambda),\ -P^0_-\leq Q\leq P^0_+,\\ \tr_{P^0_-}Q=N}}\tr_{P^0_-}\left\{ (D^0-\bar\nu\ast|\cdot|^{-1})Q\right\}.
\label{def_I}
\end{equation}
 \begin{lemma}\label{linear_BDF}
 Assume that $\bar\nu\in\cC$ is such that $\ker(D^0-\bar\nu\ast|\cdot|^{-1})=\{0\}$, and denote  the charge of the non-interacting (Furry) Dirac sea by
 $q_0:=\tr_{P^0_-}\left(\bar P-P^0_-\right)$ where $\bar P:= \chi_{(-\ii;0)}(D^0-\bar\nu\ast|\cdot|^{-1})$.
Then one has for any $N\in\Z$
 \begin{equation}
 \label{value_min_linear}
I^{\bar \nu}(N)=
 \tr_{P^0_-}\left\{\left(D^0-\bar\nu\ast|\cdot|^{-1}\right)(\bar P-P^0_-)\right\}
 +\sum_{i=1}^{|N-q_0|}|\lambda_i^\epsilon|
 \end{equation}
 where $\epsilon={\rm sgn}(N-q_0)$.
 \end{lemma}
 
 \begin{proof}
After a change of variable $Q\to Q-(\bar P-P^0_-)$, and by \cite[Lemma 1]{HLS1},
 \begin{multline}
 \label{reduc_dressed}
 I^{\bar\nu}(N)-\tr_{P^0_-}\left\{\left(D^0-\bar\nu\ast|\cdot|^{-1}\right)(\bar P-P^0_-)\right\}\\
 =\inf_{\substack{Q\in\gS^{\bar P}_1(\gH_\Lambda),\ -\bar P\leq Q\leq \bar P,\\ \tr_{\bar P}Q=N-q_0}}\tr_{\bar P}\left\{ (D^0-\bar\nu\ast|\cdot|^{-1})Q\right\}.
 \end{multline}
 By a simplified version of Proposition \ref{lieb}, one sees that the infimum of the r.h.s. of \eqref{reduc_dressed} can be restricted to states $Q$ which are a difference of two projectors: $Q=P-\bar P$ (recall $q_0\in\Z$ by \cite[Lemma 2]{HLS1}). By Theorem \ref{reduction} proved in Appendix B with $\Pi=\bar P$, there exists two orthonormal basis $(g_i)_{i=1}^{N_2}\cup (u_i)_{i\geq1}$ and $(f_i)_{i=1}^{N_1}\cup (v_i)_{i\geq1}$ respectively of $\bar P\gH_\Lambda$ and $(1-\bar P)\gH_\Lambda$, and $(l_i)\in \ell^2(\R)$ such that:
 \begin{multline}
 Q= \sum_{n=1}^{N_1}|f_n\rangle\langle f_n|-\sum_{m=1}^{N_2}|g_m\rangle\langle g_m|
 +\sum_{i\geq 1}\frac{l_i^2}{1+l_i^2}\big(|v_i\rangle\langle v_i|-|u_i\rangle\langle u_i|\big)\\
 +\sum_{i\geq 1}\frac{l_i}{1+l_i^2}\big(|u_i\rangle\langle v_i|+|v_i\rangle\langle u_i|\big)\label{expand_Q},
 \end{multline}
 with $N-q_0=N_1-N_2$. The following inequality gives the lower bound in \eqref{value_min_linear}:
 $$\tr_{\bar P}\left\{ (D^0-\bar\nu\ast|\cdot|^{-1})Q\right\}\geq \sum_{i=1}^{N_1}\lambda^+_i-\sum_{i=1}^{N_2}\lambda^-_i$$
The proof of the upper bound is left to the reader.
 \qed \end{proof}

\noindent{\bf Step 2: }{\it Upper bound.} 
To obtain an upper bound for \eqref{limit_weak_coupling}, we first fix some $\eta\geq0$ and a state $\bar P-P^0_-+\bar\gamma$, $\bar \gamma$ being a finite-rank projector commuting with $D^0-\bar\nu\ast|\cdot|^{-1}$, which satisfies
\begin{equation}
\tr(\bar\gamma)=N-q_0
\label{prop_gamma0_1}
\end{equation}
and 
\begin{equation*}
\bar\gamma\geq0\ \text{and}\ \bar\gamma\bar P=\bar P\bar\gamma=0\ \text{if}\ N-q_0>0,
\label{prop_gamma0_2}
\end{equation*}
\begin{equation}
\bar\gamma\leq0\ \text{and}\ \bar\gamma\bar P=\bar P\bar\gamma=\bar\gamma\ \text{if}\ N-q_0<0.
\label{prop_gamma0_3}
\end{equation}
We additionally assume that
\begin{equation*}
\tr_{P^0_-}\left\{ (D^0-\bar\nu\ast|\cdot|^{-1})(\bar P-P^0_-+\bar\gamma)\right\}\\
 \leq  I^{\bar\nu}(N)+\eta.
\end{equation*}
If for the considered charge $N$, then only eigenvalues appear in formula \eqref{value_min_linear}, one can of course choose $\eta=0$ and $\gamma$ to be the projector on the space spanned by any chosen eigenvectors associated with the $(\lambda_i^\epsilon)$. However, if $\lambda_i^\epsilon=\pm1$ for some $i$, then a minimizer does not necessarily exist and we can only  take an approximate one as expressed above.
Then, we recall that  \cite{LSie,HLSo}
\begin{equation}
\norm{\cD^0-D^0}_{\cB(\gH_\Lambda)}=O(\alpha), \qquad \norm{\cP^0_--P^0_-}_{\cB(\gH_\Lambda)}=O(\alpha).
\label{estim_D0}
\end{equation}
This in particular implies that the spectrum of $\cD^0-\bar\nu\ast|\cdot|^{-1}$ converges to the one of $D^0-\bar\nu\ast|\cdot|^{-1}$ as $\alpha\to0$. Therefore, $\ker(\cD^0-\bar\nu\ast|\cdot|^{-1})=\{0\}$ and
\begin{equation}
|\cD^0-\bar\nu\ast|\cdot|^{-1}|\geq \kappa/2
\label{estim_below_linear2}
\end{equation}
for $\alpha$ small enough.
We shall consider a trial state of the form
$\chi_{(-\ii;0)}(\cD^0-\bar\nu\ast|\cdot|^{-1})-\cP^0_- +\gamma_\alpha,$
where $\gamma_\alpha$ is a projector converging to and having the same rank as $\bar\gamma$. We assume moreover that it satisfies the same properties \eqref{prop_gamma0_1}, \eqref{prop_gamma0_2} and \eqref{prop_gamma0_3} as $\bar\gamma$, with $\bar P$ replaced by $\tilde P_\alpha:=\chi_{(-\ii;0)}(\cD^0-\bar\nu\ast|\cdot|^{-1})$. This is possible, since $\tilde P_\alpha-\bar P\to0$  in $\cB(\gH_\Lambda)$ as $\alpha\to0$.
 We then introduce
\begin{equation}
\tilde Q_\alpha:=\tilde P_\alpha-\cP^0_-=\chi_{(-\ii;0)}(\cD^0-\bar\nu\ast|\cdot|^{-1})-\cP^0_-
\label{def_Q_alpha}
\end{equation}
(recall that $\cD^0$ and $\cP^0_-$ implicitly depend on $\alpha$) and prove the following
\begin{lemma}\label{lemme_limit_Q_alpha}
The operator $\tilde Q_\alpha$ defined in \eqref{def_Q_alpha} satisfies:
\begin{equation}
\lim_{\alpha\to0}\norm{\tilde Q_\alpha -(\bar P-P^0_-)}_{\gS_2(\gH_\Lambda)}=0.
\label{limit_Q_alpha}
\end{equation}
Moreover, one has for $\alpha$ small enough
\begin{equation}
\tr_{\cP^0_-}(\tilde Q_\alpha)=\tr_{P^0_-}(\bar P-P^0_-)=q_0.
\label{value_trace_trial_state}
\end{equation}
\end{lemma}

\begin{proof}
Consider first the simplified case where $\bar\phi:=-\bar\nu\ast|\cdot|^{-1}\in L^2(\R^3)$. Write
\begin{align}
&\tilde Q_\alpha -(\bar P-P^0_-)  =  \frac{1}{2\pi}\int_{-\ii}^{+\ii}d\eta\frac{1}{\cD^0+i\eta}\bar\phi\frac{1}{\cD^0-\bar\nu\ast|\cdot|^{-1}+i\eta}\nonumber\\
 & \qquad\qquad\qquad\qquad\qquad - \frac{1}{2\pi}\int_{-\ii}^{+\ii}d\eta\frac{1}{D^0+i\eta}\bar\phi\frac{1}{D^0-\bar\nu\ast|\cdot|^{-1}+i\eta}\nonumber\\
  & = \frac{1}{2\pi}\int_{-\ii}^{+\ii}d\eta\frac{1}{D^0+i\eta}\bar\phi\frac{1}{\cD^0-\bar\nu\ast|\cdot|^{-1}+i\eta}(D^0-\cD^0)\frac{1}{D^0-\bar\nu\ast|\cdot|^{-1}+i\eta} \nonumber\\
& \quad+  \frac{1}{2\pi}\int_{-\ii}^{+\ii}d\eta\frac{1}{\cD^0+i\eta}(D^0-\cD^0)\frac{1}{D^0+i\eta}\bar\phi\frac{1}{\cD^0-\bar\nu\ast|\cdot|^{-1}+i\eta}.\label{Cauchy_L2}
\end{align}
Then, we use
\begin{align}
&\left\|\int_{-\ii}^{+\ii}d\eta\frac{1}{\cD^0+i\eta}(D^0-\cD^0)\frac{1}{D^0+i\eta}\bar\phi\frac{1}{\cD^0-\bar\nu\ast|\cdot|^{-1}+i\eta}\right\|_{\gS_2(\gH_\Lambda)}\nonumber\\
&\leq\norm{\frac{1}{D^0+i\eta}\bar\phi}_{\gS_2(\gH_\Lambda)}\norm{\cD^0-D^0}_{\cB(\gH_\Lambda)}\left(\int_\R  \frac{d\eta}{\sqrt{(1+\eta^2)(\kappa^2/4+\eta^2)}}\right),
\end{align}
$$\norm{\frac{1}{D^0+i\eta}\bar\phi}_{\gS_2(\gH_\Lambda)}\leq C \norm{(1+p^2)^{-1/2}}_{L^2(B(0,\Lambda))}\norm{\bar\phi}_{L^2(\R^3)}$$
by the Kato-Seiler-Simon inequality \cite{SeSi,Simon} and a similar estimate for the second term of \eqref{Cauchy_L2} to obtain
$$\norm{\tilde Q_\alpha -(\bar P-P^0_-)}_{\gS_2(\gH_\Lambda)}\leq C\norm{\cD^0-D^0}_{\cB(\gH_\Lambda)}=O(\alpha).$$

To treat the general case, it then suffices to approximate $\bar\nu$ appearing in $\tilde Q_\alpha$ and $\bar P-P^0_-$ by a function $\bar\nu_\epsilon$ such that $\bar\nu_\epsilon\ast|\cdot|^{-1}\in L^2$, uniformly with respect to $\alpha$.  This can be done by using the method of \cite{KS,HLS1}: it can be shown that there exists positive constants $C_1$ and $C_2$ independent of $\alpha$ (but which depend on $\bar\nu$) such that, for any $\nu$ satisfying $\norm{\nu-\bar\nu}_\cC\leq C_1$,
$$\norm{\tilde Q_\alpha - \left(\chi_{(-\ii;0)}\left(\cD^0-\nu\ast|\cdot|^{-1}\right)-\cP^0_-\right)}_{\gS_2(\gH_\Lambda)}\leq C_2\norm{\bar\nu-\nu}_{\cC},$$
$$\norm{(\bar P-P^0_-) - \left(\chi_{(-\ii;0)}\left(D^0-\nu\ast|\cdot|^{-1}\right)-P^0_-\right)}_{\gS_2(\gH_\Lambda)}\leq C_2\norm{\bar\nu-\nu}_{\cC}.$$
Taking for instance $\widehat{\bar\nu_\epsilon}(k)=\widehat{\bar\nu}(k)\1_{|k|\geq\epsilon}$ this allows to end the proof of \eqref{limit_Q_alpha}.

To end the proof of Lemma \ref{lemme_limit_Q_alpha}, one notices that since $\tilde Q_\alpha$ is a difference of two projectors,
\begin{equation}
\tilde Q_\alpha^2=\cP^0_+\tilde Q_\alpha\cP^0_+-\cP^0_-\tilde Q_\alpha\cP^0_-
\label{Q_alpha_diff_proj}
\end{equation}
and similarly
$(\bar P- P^0_-)^2=P^0_+(\bar P-P^0_-)P^0_+-P^0_-(\bar P-P^0_-) P^0_-.$
Therefore, \eqref{limit_Q_alpha} implies that
\begin{multline*}
\lim_{\alpha\to0}\norm{\cP^0_+\tilde Q_\alpha\cP^0_++\cP^0_-\tilde Q_\alpha\cP^0_-}_{\gS_1(\gH_\Lambda)}\\=\norm{P^0_+(\bar P-P^0_-)P^0_++P^0_-(\bar P-P^0_-) P^0_-}_{\gS_1(\gH_\Lambda)} 
\end{multline*}
from which we infer
$\lim_{\alpha\to0}\tr_{\cP^0_-}(\tilde Q_\alpha)=\tr_{P^0_-}(\bar P-P^0_-).$
Then \eqref{value_trace_trial_state} is proved since both are integers by \cite[Lemma 2]{HLS1}.
\qed \end{proof}

For $\alpha$ small enough, we deduce from \eqref{value_trace_trial_state} that
$\tr_{\cP^0_-}(\tilde Q_\alpha+\gamma_\alpha)=N$
and that $\tilde Q_\alpha+\gamma_\alpha$ is an admissible trial state.
Note that $\gamma_\alpha$ is bounded in $\gS_1(\gH_\Lambda)$, hence
$\lim_{\alpha\to 0}\E_{\rm BDF}^{\bar\nu/\alpha}(\gamma_\alpha)=\tr\{(D^0-\bar\nu\ast|\cdot|^{-1})\bar\gamma\},$
as the direct and exchange terms vanish in the limit (they are multiplied by $\alpha$).

Since $(\tilde Q_\alpha)$ satisfies \eqref{Q_alpha_diff_proj}, it is bounded in $\cQ_\Lambda$. This implies that $(\rho_{\tilde Q_\alpha})$ is uniformly bounded in $\cC$ by \eqref{continuity}, hence
\begin{equation*}
\E_{\rm BDF}^{\bar\nu/\alpha}(\tilde Q_\alpha+\gamma_\alpha)  =  \tr_{\cP^0_-}(\cD^0\tilde Q_\alpha) -D(\bar\nu,\rho_{\tilde Q_\alpha}) +\E_{\rm BDF}^{\bar\nu/\alpha}(\gamma_\alpha) + O(\alpha).
\end{equation*}
Using one more time the fact that $\tilde Q_\alpha$ is the difference of two projectors, one deduces that
$\tr_{\cP^0_-}(\cD^0\tilde Q_\alpha)=\tr(|\cD^0|\tilde Q_\alpha^2)$
which converges to
$\tr(|D^0|(\bar P-P^0_-)^2)=\tr_{P^0_-}(D^0(\bar P-P^0_-))$
by Lemma \ref{lemme_limit_Q_alpha} and \eqref{estim_D0}. Now, using \eqref{limit_Q_alpha} and the fact that $(\rho_{\tilde Q_\alpha})$ is uniformly bounded in $\cC$, we obtain
$\lim_{\alpha\to0}D(\bar\nu,\rho_{\tilde Q_\alpha})=D(\bar\nu,\rho_{\bar P-P^0_-}).$
By \cite[Lemma 5]{HLS1} which ensures
$$\tr_{P^0_-}(D^0(\bar P-P^0_-))-D(\bar\nu,\rho_{\bar P-P^0_-})=\tr_{P^0_-}\left\{\left(D^0-\bar\nu\ast|\cdot|^{-1}\right)(\bar P-P^0_-)\right\},$$
we have proved that
$$\lim_{\alpha\to0}\E_{\rm BDF}^{\bar\nu/\alpha}(\tilde Q_\alpha+\gamma_\alpha)=\tr_{P^0_-}\left\{\left(D^0-\bar\nu\ast|\cdot|^{-1}\right)(\bar P-P^0_-+\bar\gamma)\right\},$$
which means
$\limsup_{\alpha\to 0}E_{\rm BDF}^{\bar\nu/\alpha}(N)\leq I^{\bar\nu}(N)+\eta$
for any $\eta\geq0$.

\bigskip

\noindent{\bf Step 3: }{\it Lower bound.} To prove the lower bound, we consider for any fixed $\alpha$ a state $(Q_\alpha)$ satisfying
\begin{equation}
\E^{\bar\nu/\alpha}_{\rm BDF}(Q_\alpha)\leq E^{\bar\nu/\alpha}(N) +\alpha,\qquad \tr_{\cP^0_-}(Q_\alpha)=N.
\label{approx_min_lower_bound}
\end{equation}
By Proposition \ref{lieb}, we may moreover assume that $Q_\alpha=P_\alpha-\cP^0_-$ where $P_\alpha$ is an orthogonal projector. 
Let us show that the sequence $(Q_\alpha)$ is bounded in $\gS_2(\gH_\Lambda)$. To this end, we first notice that $E^{\bar\nu/\alpha}(N)$ is bounded from above by the previous step and therefore, by \eqref{approx_min_lower_bound} and Kato's inequality for the exchange term, we obtain the bound
$$(1-\alpha\pi/4)\tr(|\cD^0|Q_\alpha^2)\leq C+ \frac{1}{2\alpha} D(\bar\nu,\bar\nu),$$
which proves that $(\sqrt{\alpha}Q_\alpha)$ is bounded in $\gS_2(\gH_\Lambda)$. By Kato's inequality, this means that the exchange term satisfies
$$\frac\alpha2\iint_{\R^6}\frac{|Q_\alpha(x,y)|^2}{|x-y|}dx\,dy\leq \frac{\alpha\pi}{4}\tr(|\cD^0|Q_\alpha^2)\leq \frac{\pi (\alpha C+ D(\bar\nu,\bar\nu)/2)}{4(1-\alpha\pi/4)}$$
and it is thus uniformly bounded in $\alpha$. Therefore, one has
\begin{equation}
\tr_{\cP^0_-}(\cD^0Q_\alpha)-D(\bar\nu,\rho_{Q_\alpha})+\frac\alpha2 D(\rho_{Q_\alpha},\rho_{Q_\alpha})\leq C
\label{bound_eq_lower}
\end{equation}
for some other constant $C$ independent of $\alpha$. By \cite[Lemma 5]{HLS1} we have
\begin{eqnarray}
\tr_{\cP^0_-}(\cD^0Q_\alpha)-D(\bar\nu,\rho_{Q_\alpha}) & = & \tr_{\cP^0_-}\{(\cD^0-\bar\nu\ast|\cdot|^{-1})Q_\alpha\}\label{calcul_lower_bd_1}\\
 & = & \tr_{\tilde P_\alpha}\{(\cD^0-\bar\nu\ast|\cdot|^{-1})(Q_\alpha-\tilde Q_\alpha)\}\nonumber\\ 
  & & \qquad  +\tr_{\cP^0_-}\{(\cD^0-\bar\nu\ast|\cdot|^{-1})\tilde Q_\alpha\}\label{calcul_lower_bd_2}\\
   & \geq & \label{calcul_lower_bd_3} \tr\{|\cD^0-\bar\nu\ast|\cdot|^{-1}|(Q_\alpha-\tilde Q_\alpha)^2\} -C.
\end{eqnarray}
In \eqref{calcul_lower_bd_2}, we have inserted 
$\tilde Q_\alpha=\chi_{(-\ii;0)}(\cD^0-\bar\nu\ast|\cdot|^{-1})-\cP^0_-= \tilde P_\alpha -\cP^0_-$
used in Step 1, and we have applied \cite[Lemma 1]{HLS1} allowing to change the reference projector in the trace. In \eqref{calcul_lower_bd_3}, we have used the fact that $Q_\alpha-\tilde Q_\alpha$ satisfies
$-\tilde P_\alpha\leq Q_\alpha-\tilde Q_\alpha\leq 1-\tilde P_\alpha.$
We have also used that $\tr_{\cP^0_-}\{(\cD^0-\bar\nu\ast|\cdot|^{-1})\tilde Q_\alpha\}$ is uniformly bounded since it converges to $\tr_{P^0_-}\{(D^0-\bar\nu\ast|\cdot|^{-1})(\bar P- P^0_-)\}$, as proved in the first step.
By \eqref{bound_eq_lower} and \eqref{calcul_lower_bd_3}, we infer that 
$$\tr\{|\cD^0-\bar\nu\ast|\cdot|^{-1}|(Q_\alpha-\tilde Q_\alpha)^2\}+\frac\alpha2 D(\rho_{Q_\alpha},\rho_{Q_\alpha})\leq C$$
for some uniform constant $C$. 
Using \eqref{estim_below_linear2} and Lemma \ref{lemme_limit_Q_alpha}, we eventually deduce that $(Q_\alpha)$ is bounded in $\gS_2(\gH_\Lambda)$.
Now, we can write
\begin{eqnarray}
\E_{\rm BDF}^{\bar\nu/\alpha}(Q_\alpha) & \geq &  \tr_{\cP^0_-}\{(\cD^0-\bar\nu\ast|\cdot|^{-1})Q_\alpha\}+O(\alpha)\\
 & \geq & \inf_{\substack{Q\in\cQ_\Lambda\\ \tr_{\cP^0_-}(Q)=N}}\tr_{\cP^0_-}\{(\cD^0-\bar\nu\ast|\cdot|^{-1})Q\}+O(\alpha)\label{estim_below_inf}
\end{eqnarray}
since the exchange term is $O(\alpha)$ and the direct term is $\geq0$. Next we have
$$\lim_{\alpha\to0}\inf_{\substack{Q\in\cQ_\Lambda\\ \tr_{\cP^0_-}(Q)=N}}\tr_{\cP^0_-}\{(\cD^0-\bar\nu\ast|\cdot|^{-1})Q\}= I^{\bar\nu}(N)$$
defined in \eqref{def_I}. It suffices to compute the above infimum by a formula similar to \eqref{value_min_linear} and to use the convergence of the spectrum of $\cD^0-\bar\nu\ast|\cdot|^{-1}$ to the one of $D^0-\bar\nu\ast|\cdot|^{-1}$.
This shows that
$\liminf_{\alpha\to 0}E_{\rm BDF}^{\bar\nu/\alpha}(N)\geq I^{\bar\nu}(N)$
which ends the proof of \eqref{limit_weak_coupling}.

\bigskip

\noindent{\bf Step 4:} {\it Existence of a minimizer for $\alpha$ small enough.} We now assume that $\bar\nu$ satisfies the assumptions $(a)$ and $(b)$ of Theorem \ref{exists_weak_limit}. Since $D^0-t\bar\nu\ast|\cdot|^{-1}$ has no eigenvalue which crosses 0 when $t\in[0;1]$, one classically deduces that 
$q_0=\tr_{P^0_-}(\bar P-P^0_-)=0.$
Hence,
$I^{\bar\nu}(N)=\sum_{i=1}^N\lambda^+_i$ when $N\geq0.$
On the other hand, $I^{0}(N)=|N|$ for all $N\in\Z$.
This shows that $(H_1')$ is satisfied for the noninteracting linear model obtained in the limit $\alpha\to0$. To prove that $(H_1')$ holds for $\alpha$ small enough is not difficult. We just have to prove that only finitely many strict inequalities have to be checked in $(H_1')$, and then to apply \eqref{limit_weak_coupling}. Unfortunately, we cannot use Lemma \ref{encadrement_min_prop} since the lower bound of \eqref{encadrement_min} diverges when $\nu=\bar\nu/\alpha$. Instead, we prove the following
\begin{lemma}\label{infini_unif} We assume that $\bar\nu\in\cC$ is such that ${\rm ker}(D^0-\bar\nu\ast|\cdot|^{-1})=\{0\}$. Then there exists $0<\alpha_0<4/\pi$ and positive constants $\kappa_1,\kappa_2$ such that, for any $\alpha\in [0;\alpha_0]$,
$E^{\bar\nu/\alpha}(N) \geq \kappa_1|N|-\kappa_2.$
Therefore, there exists a positive constant $K_0$ independent of $\alpha$ such that $E^{\bar\nu/\alpha}(N)<E^{\bar\nu/\alpha}(N-K)+E^{0}(K)$ for all $|K|\geq K_0$.
\end{lemma}

\begin{proof}
We argue like in Step 3. Let $Q\in\cQ_\Lambda$ be a state such that 
$\E_{\rm BDF}^{\bar\nu/\alpha}(Q)\leq E^{\bar\nu/\alpha}(N)+\eta$ and $\tr_{\cP^0_-}(Q)=N$
for some fixed $\eta>0$. Then by \eqref{encadrement_min},
$$(1-\alpha\pi/4)\tr(|\cD^0|Q^2)\leq g_0(0)|N|+\frac{1}{2\alpha}D(\bar\nu,\bar\nu)+\eta$$
and therefore
$$\frac{\alpha}{2}\iint_{\R^6}\frac{|Q(x,y)|^2}{|x-y|}dx\,dy\leq \frac{\pi}{4(1-\alpha\pi/4)}(g_0(0)\alpha|N|+D(\bar\nu,\bar\nu)/2+\eta\alpha).$$
We obtain
\begin{multline*}
 \E_{\rm BDF}^{\bar\nu/\alpha}(Q)\geq \tr_{\cP^0_-}(\cD^0Q)-D(\bar\nu,\rho_Q)\\-\frac{\pi}{4(1-\alpha\pi/4)}(g_0(0)\alpha|N|+D(\bar\nu,\bar\nu)/2+\eta\alpha).
\end{multline*}
For $\alpha$ small enough, we have
\begin{align*}
&\tr_{\cP^0_-}(\cD^0Q)-D(\bar\nu,\rho_Q) \\
&\qquad = \tr_{\cP^0_-}((\cD^0-\nu\ast|\cdot|^{-1})Q)\\
 &\qquad = \tr_{\tilde P_\alpha}((\cD^0-\nu\ast|\cdot|^{-1})(Q+\tilde Q_\alpha)) -\tr_{\cP^0_-}((\cD^0-\nu\ast|\cdot|^{-1})\tilde Q_\alpha)\\
 & \qquad\geq (\kappa/2)|\tr_{\cP^0_-}(Q+\tilde Q_\alpha)|- \tr_{\cP^0_-}((\cD^0-\nu\ast|\cdot|^{-1})\tilde Q_\alpha)\\
 & \qquad\geq \kappa|N|/2-\kappa|\tr_{\cP^0_-}(\tilde Q_\alpha)|/2- \tr_{\cP^0_-}((\cD^0-\nu\ast|\cdot|^{-1})\tilde Q_\alpha)
\end{align*}
(recall that $\tilde Q_\alpha$ is defined in \eqref{def_Q_alpha}). We deduce that
$$E^{\bar\nu/\alpha}(N)\geq |N|\left( \kappa/2-\alpha\frac{\pi g_0(0)}{4(1-\alpha\pi/4)}\right)-\kappa_2$$
for some uniform constant $\kappa_2$. Recall that $g_0(0)$ implicitly depends on $\alpha$, and that it converges to 1 as $\alpha\to0$, see \cite{LSie,HLSo}.
\qed \end{proof}

Once we know that there exists a minimizer $Q_\alpha$ of $E^{\bar\nu/\alpha}(N)$, it is an easy adaptation of the previous arguments to prove that $Q_\alpha$ takes the form \eqref{form_Q_weak_coupling} and behaves as stated. This ends the proof of Theorem \ref{exists_weak_limit}. \qed

\section{Proof of Theorem \ref{non-rel}}

\noindent\textbf{Step 1:} \textit{Scaling properties and the spectrum of $\cD^0$ when $c\gg1$.} To avoid any confusion, we shall use the following notation in the proof: we denote by $E^\nu_{\alpha,c,\Lambda}(N)$ the infimum of the BDF energy in $\cQ_\Lambda(N)$, depending on the coupling constant $\alpha$, the speed of light $c$ and the ultraviolet cut-off $\Lambda$. We are then interested in the limit of $E^\nu_{1,c,\Lambda_0c}$ as $c\to\ii$.
Since most of our previous results are expressed in terms of the coupling constant $\alpha=1/c$, we shall often use in this proof the following obvious scaling property
\begin{equation}
E^\nu_{1,c,\Lambda_0 c}(N)=c^2 E^{\nu_c}_{1/c,1,\Lambda_0}(N),
\label{scaling}
\end{equation}
with $\nu_c(x)=c^{-3}\nu(x/c)$.
More precisely, we introduce the following operator 
\begin{equation}
\begin{array}{cccc}
U_c:= & \gH_{c\Lambda_0} & \longmapsto  & \gH_{\Lambda_0}\\
 & \psi & \longrightarrow &  (U_c\psi)(x)=c^{-3/2}\psi(x/c).
\end{array}
\label{scaling_op}
\end{equation}
Then for any state $Q\in\cQ_{c\Lambda_0}(N)$, $\tilde Q_c:=U_cQU_c^*$ belongs to $\cQ_{\Lambda_0}(N)$, and 
$\E^{\nu}_{1,c,c\Lambda_0}(Q)=c^2\E^{\nu_c}_{1/c,1,\Lambda_0}(U_cQU_c^*).$
To avoid both any confusion and any complicated notation, we shall always denote by $\cD^0$ and $\cP^0_-$ the free mean-field operator and the free projector when $\alpha=1$, $\Lambda=c\Lambda_0$ and the speed of light is $c$. For the other equivalent units where $\alpha=1/c$, we use the following notation:
$$\tilde\cP^0_-:=U_c\cP^0_-U_c^* \quad \text{and}\quad \tilde\cD^0=\frac{U_c\cD^0U_c^*}{c^2}.$$
It will be implicit below that $\cD^0$, $\tilde\cD^0$, $\cP^0_-$ and $\tilde\cP^0_-$ indeed all depend on $c$ and $\Lambda_0$. We similarly write 
$$\cD^0(p)=g_1(|p|)\alp\cdot\omega_p+g_0(|p|)\beta,\quad \tilde\cD^0(p)=\tilde g_1(|p|)\alp\cdot\omega_p+\tilde g_0(|p|)\beta$$
and notice that $\tilde g_0$ and $\tilde g_1$ are uniformly bounded with respect to $\alpha=1/c$ and satisfy
$g_0(x)=c^2\tilde g_0(x/c)$ and $g_1(x)=c^2\tilde g_1(x/c)$.
For $c$ large enough, we are able to identify the essential spectrum threshold of $\cD^0$ as stated in the following
\begin{lemma}\label{threshold}
Assume that $c$ is large enough, then there exist  $\kappa,\kappa'>0$ depending only on $\Lambda_0$ such that
\begin{equation}
\sqrt{g_0(0)^2+\kappa c^2|p|^2}\leq |\cD^0(p)|\leq g_0(0)+(1+{\kappa'}/{c})\frac{|p|^2}{2}
\label{estim_D0_c}
\end{equation}
for any $p\in B(0,c\Lambda_0)$. In particular, $\min\sigma(|\cD^0|)=g_0(0)$.
\end{lemma}

\begin{proof}
It is known that $p\mapsto\cD^0(p)$ is smooth \cite{HLSo}, hence $p\mapsto g_0(|p|)$ and $p\mapsto g_1(|p|)/|p|$ are also smooth. By \cite{LS,HLSo}, there exist two continuous functions $h_0$ and $h_1$, uniformly bounded on $B(0,\Lambda_0)$ with respect to $c$, such that
\begin{equation}
\tilde g_0(|p|)=\tilde g_0(0)+\frac{|p|^2}{c}h_0(|p|),\quad
\tilde g_1(|p|)=|p|(\tilde g_1'(0)+\frac{|p|^2}{c} h_1(|p|)),
\label{prop_g_alpha}
\end{equation}
where  $\tilde g_0(0)=1+O(1/c)$ and $\tilde g_1'(0)=1+O(1/c)$.
We infer from \eqref{prop_g_alpha} that 
\begin{equation}
g_0(|p|)=c^2\tilde g_0(0)+\frac{|p|^2}{c}h_0(|p|/c),\quad
g_1(|p|)=c|p|(\tilde g_1'(0)+ \frac{|p|^2}{c^3}h_1(|p|/c)).
\label{prop_g}
\end{equation}
Therefore there exist two positive constants $\kappa_0$ and $\kappa_1$ such that
$g_0(|p|)^2\geq c^2\tilde g_0(0)^2-\kappa_0c|p|^2$ and $g_1(|p|)^2\geq \kappa_1c^2|p|^2$, 
hence
$$g_0(|p|)^2+g_1(|p|)^2\geq c^2\tilde g_0(0)^2+c^{2}|p|^2(\kappa_1-\kappa_0/c)\geq g_0(0)^2+\kappa c^{2}|p|^2$$
for $c$ large enough.
Similarly, we notice that, by \cite[Theorem 2.2]{HLSo}
\begin{eqnarray*}
 \sqrt{g_0(|p|)^2+g_1(|p|)^2}-g_0(0)& \leq & g_0(|p|)\sqrt{1+|p|^2/c^2}-g_0(0)\\
 & \leq & g_0(0)\frac{|p|^2}{2c^2}+\frac{|p|^2}{c}\norm{h_0}_{L^\ii}\sqrt{1+\Lambda_0^2},
\end{eqnarray*}
which ends the proof of \eqref{estim_D0_c} since $g_0(0)/c^2=\tilde g_0(0)=1+O(1/c)$.
\qed \end{proof}

\bigskip

\noindent\textbf{Step 2:} \textit{Upper bound}. Let us start by proving the upper bound
\begin{equation}
\limsup_{c\to\ii} \{E^\nu_{1,c,c\Lambda_0}(N)-N\,g_0(0)\}\leq E_{\rm HF}^\nu(N).
\label{limit_sup}
\end{equation}
Let $c_n\to\ii$ be a sequence which realizes the $\limsup$ in \eqref{limit_sup}. Let $\psi=(\psi_1,...,\psi_N)$ be a minimizer of the Hartree-Fock energy \cite{LS,Lions-87}, belonging to $H^2(\R^3,\C^2)^N$. We introduce the following subspace of $\gH_+^0$
$$W_n:={\rm Span}\left\{\cP^0_+\phi_i, i=1,...,N\right\},\qquad \phi_i=\left(\begin{array}{c}
\psi_i\\ 0
\end{array}\right).$$
By \eqref{prop_g}, $\pscal{\cP^0_+\phi_i,\cP^0_+\phi_j}=\delta_{ij}+O(1/c_n)$ and we can thus choose an orthonormal basis $(\phi_1^n,...,\phi_N^n)$ of $W_n$ which satisfies $\norm{\phi_i^n-\phi_i}_{H^1}\to0$ as $c_n\to\ii$.
We then take
$\gamma^n=\sum_{i=1}^N|\phi_i^n\rangle\langle\phi_i^n|$
as a trial state. Using \eqref{estim_D0_c}, we infer
\begin{eqnarray*}
\tr_{\cP^0_-}(\cD^0\gamma^n)-N\,g_0(0) & = & \sum_{i=1}^N\pscal{\left(\sqrt{g_0(|p|)^2+g_1(|p|)^2}-g_0(0)\right)\phi_i^n,\phi_i^n}\\
 & \leq & \frac{1+\kappa'/c}{2}\sum_{i=1}^N\pscal{|p|^2\phi_i^n,\phi_i^n}.
\end{eqnarray*}
This allows to prove the upper bound \eqref{limit_sup}.

\bigskip

\noindent\textbf{Step 3:} \textit{Lower bound: construction of an approximate solution}. The main part of the proof will now consist in showing the lower bound
\begin{equation}
\liminf_{c\to\ii} \{E^\nu_{1,c,c\Lambda_0}(N)-N\,g_0(0)\}\geq E_{\rm HF}^\nu(N),
\label{limit_inf}
\end{equation}
which will end the proof of \eqref{limit_energy_nonrel}. 
To this end, we consider a sequence $c_n\to\ii$ which realizes the $\liminf$ in \eqref{limit_inf}.
For any $c_n$, we shall need a state $Q_n$ which is not only an approximate minimizer of $E^\nu_{1,c_n,c_n\Lambda_0}(N)$, but also an approximate solution of the self-consistent equation. Such a state will be obtained by a general perturbation result due to Borwein and Preis \cite{BP} (see also \cite{Gho}), and which we state in the simplified Hilbert case as follows:

\begin{theorem}[A smooth variational perturbation principle \cite{BP,Gho}]\label{perturbation_principle} Let $\mathcal{M}$ be a closed subset of a Hilbert space $\mathcal{H}$, and $F:\mathcal{M}\mapsto (-\ii;+\ii]$ be a lower semi-continuous function that is bounded from below and not identical to $+\ii$. For all $\epsilon>0$ and all $u\in \mathcal{M}$ such that $F(u)< \inf_\mathcal{M}F+\epsilon^2$, there exist $v\in \mathcal{M}$ and $w\in \overline{{\rm Conv}(\mathcal{M})}$ such that 
\begin{enumerate}
\item $F(v)<\inf_\mathcal{M}F+\epsilon^2$
\item $\norm{u-v}_\mathcal{H}<\sqrt{\epsilon},\ \ \norm{v-w}_\mathcal{H}<\sqrt{\epsilon}$
\item $F(v)+\epsilon\norm{v-w}_\mathcal{H}^2=\min\left\{F(z)+\epsilon\norm{z-w}_\mathcal{H}^2,\ z\in \mathcal{M}\right\}$.
\end{enumerate} 
\end{theorem} 

We apply this result by taking $F=\E_{1,c_n,c_n\Lambda_0}^\nu$, $\mathcal{H}=\gS_2(\gH_{c_n\Lambda_0})$, $\epsilon=c_n^{-2}$ and
\begin{align*}
 \mathcal{M}&:=\left\{P-\cP^0_-\in\gS_2(\gH_{c_n\Lambda_0})\ |\ P=P^2=P^*,\  \tr_{\cP^0_-}(P-\cP^0_-)=N\right\}\\
&\subset \cQ_{c_n\Lambda_0}(N).
\end{align*}
We recall that the BDF functional $\E_{1,c_n,c_n\Lambda_0}^\nu$ is continuous on $\cQ_{c_n\Lambda_0}(N)$ for the  $\gS_1^{\cP^0_-}(\gH_{c_n\Lambda_0})$ topology, hence on $\mathcal{M}$ for the $\gS_2(\gH_{c_n\Lambda_0})$ topology. One has
$E_{1,c_n,c_n\Lambda_0}^\nu(N)=\inf_{\mathcal{M}}\E_{1,c_n,c_n\Lambda_0}^\nu$
by Lieb's variational principle, Proposition \ref{lieb}. Notice also that $\overline{{\rm Conv}\mathcal{M}}=\cQ_{c_n\Lambda_0}(N)$.
Applying Theorem \ref{perturbation_principle}, we therefore obtain an orthogonal projector $P^n$ on $\gH_{c_n\Lambda_0}$ and a state $R^n\in \cQ_{c_n\Lambda_0}(N)$ such that
$Q^n:=P^n-\cP^0_-\in\mathcal{M}\subset \cQ_{c_n\Lambda_0}(N)$ minimizes the following perturbed functional
$Q\in\mathcal{M}\mapsto \E_{1,c_n,c_n\Lambda_0}^\nu(Q)+\frac1{c_n^2}\tr\{(Q-R^n)^2\}$
on $\mathcal{M}$ and satisfies
$$\E_{1,c_n,c_n\Lambda_0}^\nu(Q^n)\leq E_{1,c_n,c_n\Lambda_0}^\nu(N)+c_n^{-4},\qquad \norm{Q^n-R^n}_{\gS_2(\gH_{c_n\Lambda_0})}\leq c_n^{-1}.$$
Noticing that
$$\tr\{(Q^n-R^n)^2\}=2\tr_{\cP^0_-}\left\{Q^n(1/2-\cP^0_--R^n)\right\}+\tr((R^n)^2),$$
since $\tr((Q^n)^2)=\tr((Q^n)^{++}-(Q^n)^{--})=2\tr_{\cP^0_-}\{Q^n(1/2-\cP^0_-)\}$, it is then an easy adaptation of Proposition \ref{prop_scf} to prove that $Q^n$ satisfies the following equation, for some $\mu^n\in\R$,
\begin{equation}
\label{approx_sol}
Q^n+\cP^0_-=P^n=\chi_{(-\ii;\mu^n]}\left(\cD_{Q^n}+\frac{2}{c_n^2}(1/2-\cP^0_--R^n)\right).
\end{equation}
We then introduce the approximate vacuum solution
$$Q_{\rm vac}^n:=\chi_{(-\ii;0)}\left(\cD_{Q^n}+\frac{2}{c_n^2}(1/2-\cP^0_--R^n)\right)-\cP^0_-$$
and the approximate electronic solution $\gamma^n:=Q^n-Q_{\rm vac}^n$.

\bigskip

\noindent\textbf{Step 3:} \textit{Estimates on the approximate vacuum solution $Q_{\rm vac}^n$}. To apply previous results, we now introduce $\tilde Q^n:= U_{c_n}Q^nU_{c_n}^*$, $\tilde Q_{\rm vac}^n:= U_{c_n}Q_{\rm vac}^nU_{c_n}^*$
where $U_{c_n}$ is the scaling operator defined above in \eqref{scaling_op}. One has
\begin{multline}
\tilde Q_{\rm vac}^n=\chi_{(-\ii;0)}\left(\tilde\cD^0+c_n^{-1}(\rho_{\tilde Q^n}-\nu_{c_n})\ast|\cdot|^{-1}-c_n^{-1}\frac{\tilde Q^n(x,y)}{|x-y|}\right.\\\left.-\frac{2}{c_n^4}\tilde R^n+\frac{2}{c_n^4}(1/2-\tilde\cP^0_-)\right)-\tilde\cP^0_-.
\label{approx_eq_Gamma}
\end{multline}
Notice the obvious property $\tilde\cP^0_-=\chi_{(-\ii;0)}\left(\tilde\cD^0+\frac{2}{c_n^4}(1/2-\tilde\cP^0_-)\right)$
for $n$ large enough, and that $D(\nu_{c_n},\nu_{c_n})=c_n^{-1}D(\nu,\nu)$.
Since $\tilde Q^n$ satisfies
$$E_{1/c_n,1,\Lambda_0}(N)\leq \E_{1/c_n,1,\Lambda_0}(\tilde Q^n)\leq E_{1/c_n,1,\Lambda_0}(N)+c_n^{-6},$$
$(\tilde Q^n)_{n\geq1}$ is bounded uniformly in $\gS_1^{\cP^0_-}(\gH_{\Lambda_0})$, for $n$ large enough by Lemma \ref{lemma_continuity}. In particular, $\rho_{\tilde Q^n}$ is uniformly bounded in $\cC$ for $n$ large enough. We deduce from the equation \eqref{approx_eq_Gamma} satisfied by $\tilde Q_{\rm vac}^n$ and the results of \cite{HLS1,HLS2} that
$$\tr\left((1+|\nabla|)(\tilde Q_{\rm vac}^n)^2\right)^{1/2}+\norm{\rho_{\tilde Q_{\rm vac}^n}}_\cC=O(1/c_n).$$
Using now
$$\tr((Q_{\rm vac}^n)^2)=\tr((\tilde Q_{\rm vac}^n)^2)\ \text{and}\ D(\rho_{Q_{\rm vac}^n},\rho_{Q_{\rm vac}^n})=c_n\,D(\rho_{\tilde Q_{\rm vac}^n},\rho_{\tilde Q_{\rm vac}^n}),$$
we eventually obtain
\begin{equation}
\norm{Q_{\rm vac}^n}_{\gS_2(\gH_{c_n\Lambda_0})}=O(1/c_n)\ \text{and}\  \tr\left(|\nabla|(Q^n_{\rm vac})^2\right)^{1/2}+\norm{\rho_{Q_{\rm vac}^n}}_\cC=O(c_n^{-1/2}).
\label{estim_vacuum}
\end{equation}

\bigskip

\noindent\textbf{Step 4:} \textit{Non-relativistic limit of the approximate electronic solution $\gamma^n$ and proof of the lower bound \eqref{limit_inf}}. By \eqref{estim_vacuum}, we have $\norm{P^n-\cP^0_-}_{\cB(\gH_{c_n\Lambda_0})}<1$ for $n$ large enough, and therefore that the vacuum has a vanishing charge \cite[Lemma 2]{HLS1}:
$\tr_{\cP^0_-}(Q_{\rm vac}^n)=0.$
Since by construction the full state $Q^n$ has a total charge $N>0$, this means that necessarily $\mu^n$ in \eqref{approx_sol} is a positive real constant, and that the perturbed mean-field operator 
$\cD_{Q^n}+2(1/2-\cP^0_--R^n)/c_n^2$
has at least $N$ positive eigenvalues. The operator $\gamma^n$ is then the projector on the $N$ first positive eigenstates: we can write
$\gamma^n=\sum_{i=1}^N|\phi_i^n\rangle\langle\phi^n_i|,$
where each $\phi_i^n$ is a solution of the following equation
\begin{equation}
\left(\cD_{Q^n}+\frac{2}{c_n^2}(1/2-\cP^0_--R^n)\right)\phi_i^n=\mu_i^n\phi_i^n,
\label{eq_orbitales}
\end{equation}
$(\mu_i^n)_{i=1}^N$ being the $N$ first positive eigenvalues of $\cD_{Q^n}+\frac{2}{c_n^2}(1/2-\cP^0_--R^n)$. In order to prove the lower bound \eqref{limit_inf}, we shall now show that $\Phi^n=(\phi_1^n,...,\phi_N^n)$ converges to a solution of the Hartree-Fock equations. To this end, we use ideas from Esteban and Séré \cite{ES2}: we prove that $(\Phi^n)$ is bounded in $H^1(\R^3,\C^4)^N$ and that each $\mu_i^n$ stays away from the essential spectrum of $\cD_{Q^n}+\frac{2}{c_n^2}(1/2-\cP^0_--R^n)$ as $n$ grows. We then apply a result of Lions \cite{Lions-87}.
\begin{lemma}\label{estim_vp_above} There exists a constant $\epsilon>0$ depending only on $N$, $\nu$ and $\Lambda_0$, such that
\begin{equation}
\forall i=1,...,N,\qquad \limsup_{n\to\ii}(\mu_i^n-g_0(0))\leq -\epsilon<0.
\end{equation}
\end{lemma}

\begin{proof}
First, we notice that for any $i=1,...,N$, $\mu_i^n$ is at most the $N$th eigenvalue of the following operator, with the self-interaction removed:
$$\cD_i^n:=\cD^0+\rho^n_i\ast|\cdot|^{-1}-\frac{\gamma^n_i(x,y)}{|x-y|}+\cK_n$$
with $\rho_i^n=\sum_{j\neq i}|\phi_j^n|^2-\nu$, $\gamma_i^n(x,y)=\sum_{j\neq i}\phi_j^n(x)\phi_j^n(y)^*$,
$$\cK_n=\rho_{Q_{\rm vac}^n}\ast|\cdot|^{-1}-\frac{Q_{\rm vac}^n(x,y)}{|x-y|}+\frac{2}{c_n}(1/2-\cP^0_--R^n).$$
We estimate $\mu_N^n$ by means of the min-max characterization of the eigenvalues in the gap which was proposed by Dolbeault, Esteban and Séré in \cite{DES}. By the continuation principle of \cite{DES}, one can prove that the assumptions of \cite[Theorem 1]{DES} are satisfied for $c_n$ large enough, and therefore the first $N$ eigenvalues of $\cD_i^n$ are given by the formula
\begin{equation}
\mu_k(\cD_i^n)=\inf_{\substack{V\subset \gH^0_+\\ \dim V=k}}\sup_ {\substack{\phi\in V\oplus \gH^0_-\\ \norm{\phi}_{L^2}=1}}\pscal{\cD_i^n\phi,\phi},\qquad k=1,...,N.
\label{minmaxvp}
\end{equation}

We then argue like in \cite[Lemma 4.5]{ES1} to estimate $\mu_N(\cD_i^n)\geq \mu_i^n$. We choose an $N$-dimensional vector subspace $W$ of $H^1(\R^3,\R)$, of smooth radial functions with compact support in the Fourier domain, and introduce
$$V_R:=\left\{x\mapsto R^{-3/2}\left(\begin{matrix}f(x/R)\\0\\0\\0\\ \end{matrix}\right),\ f\in W\right\}.$$
It is clear that $\cP^0_+V_R$ is an $N$-dimensional vector space for $c_n$ large enough, uniformly in $R\geq1$. We then use $\gH^0_-\oplus\cP^0_+V_R=\gH^0_-+ V_R$ to estimate $\mu_N(\cD_i^n)$ by Formula \eqref{minmaxvp}. Let be $\phi\in\gH^0_-+ V_R$ such that $\phi=\phi_-+\chi$ with $\phi_-\in \gH_-^0$ and $\chi\in V_R$. We compute
\begin{equation*}
\pscal{\cD_i^n\phi,\phi} = \pscal{\cD_i^n\phi_-,\phi_-}+\pscal{\cD_i^n\chi,\chi}+2\Re(\pscal{\cD_i^n\phi_-,\chi}).
\end{equation*}
First, we use Kato's inequality to obtain
\begin{align*}
\left|\iint\frac{\rho_i^n(x)|\phi_-(y)|^2}{|x-y|}dx\,dy\right|&\leq \frac{\pi}{2}\norm{\rho_i^n}_{L^1}\pscal{|\nabla|\phi_-,\phi_-}\\
&\leq \frac{\pi(Z+N-1)}{2c_n}\pscal{|\cD^0|\phi_-,\phi_-}. 
\end{align*}
The same argument with Hardy's inequality leads to
\begin{align*}
\left|\iint\frac{\rho_i^n(x)\phi_-(y)\chi(x)}{|x-y|}dx\,dy\right|&\leq2(Z+N-1)\norm{\phi_-}_{L^2}\norm{\nabla\chi}_{L^2}\\
&\leq \norm{\phi_-}_{L^2}^2+\frac{\kappa_1}{R^2}\norm{\chi}_{L^2}^2 
\end{align*}
for some constant $\kappa_1>0$ independent of $c_n$. Similarly, we write
$$|\pscal{\cD^0\phi_-,\chi}|\leq \norm{|\cD^0|^{1/2}\phi_-}_{L^2}\norm{\cP^0_-|\cD^0|^{1/2}\chi}_{L^2}.$$
Then, we notice that
$$\norm{\cP^0_-|\cD^0|^{1/2}\chi}_{L^2}^2=\pscal{|\cD^0|\cP^0_-\chi,\chi}=\pscal{\left\{\sqrt{g_0(|p|)^2+g_1(|p|)^2}-g_0(|p|)\right\}\chi,\chi}/2.$$
By Lemma \ref{threshold}
\begin{equation}
\sqrt{g_0(|p|)^2+g_1(|p|)^2}-g_0(|p|)\leq 2\kappa_2|p|^2
\label{estim_nonrel_limit}
\end{equation}
for some constant $\kappa_2>0$ depending only on $\Lambda_0$ and for $c_n$ large enough, which proves that
$$|\pscal{\cD^0\phi_-,\chi}|\leq \kappa_2\pscal{|\cD^0|\phi_-,\phi_-}^{1/2}\norm{\nabla\chi}_{L^2}\leq  \frac{\pscal{|\cD^0|\phi_-,\phi_-}}{4}+\frac{\kappa_3}{R^2}\norm{\chi}_{L^2}^2.$$
We now estimate the term $\pscal{\cD_i\chi,\chi}$. First we use \eqref{prop_g} to obtain
$$\pscal{\cD^0\chi,\chi}=\pscal{g_0(|p|)\chi,\chi}\leq g_0(0)\norm{\chi}_{L^2}^2+\frac{\kappa_4}{c_nR^2}\norm{\chi}^2_{L^2}.$$
Then, we write
$$\iint_{\R^6}\frac{\rho_i^n(x)\chi(y)^2}{|x-y|}dx\,dy\leq (N-1)\int_{\R^3}\frac{\chi(y)^2}{|y|}dy$$
where we have used that $\chi$ is a radial function. On the other hand, $\nu$ being fixed in $L^1$, one has
$$\iint_{\R^6}\frac{\nu(x)\chi(y)^2}{|x-y|}dx\,dy\geq Z\int_{\R^3}\frac{\chi(y)^2}{|y|}dy - o(R^{-1}\norm{\chi}_{L^2}^2).$$
Eventually, we estimate the term involving $\cK_n$. We use
$$\norm{1/2-\cP^0_-+R^n}_{\cB(\gH_{c_n\Lambda_0})}=O_{c_n\to\ii}(1),$$
$$\norm{\rho_{Q_{\rm vac}^n}\ast|\cdot|\phi}_{L^2}\leq \norm{\rho_{Q_{\rm vac}^n}\ast|\cdot|}_{L^6}\norm{\phi}_{L^3}\leq \kappa_5D(\rho_{Q_{\rm vac}^n},\rho_{Q_{\rm vac}^n})^{1/2}\norm{\phi}_{H^1(\R^3)}$$
and a similar inequality for $Q^n_{\rm vac}(x,y)/|x-y|$ to prove that for some constant $\kappa_6>0$,
$|\cK_n|^2\leq (\kappa_6)^2(1-\Delta)/c_n$, and therefore
$|\cK_n|\leq \kappa_6 c_n^{-1/2}\sqrt{1-\Delta}.$
Using the same method as above to estimate the term $\pscal{\cK_n\phi,\phi}$ and the fact that $\gamma_i^n(x,y)/|x-y|$ defines a nonnegative operator, we therefore obtain the bound
\begin{eqnarray*}
\pscal{\cD_i^n\phi,\phi} & \leq &  \left(-1/2+O(c_n^{-1/2})\right)\pscal{|\cD^0|\phi_-,\phi_-}+2\norm{\phi_-}_{L^2}^2\\
 & & +\left(g_0(0)+O(c_n^{-1/2})+\frac{\kappa_7(N-1-Z)}{R}+o(R^{-1})\right)\norm{\chi}_{L^2}^2.
\end{eqnarray*}
Finally, by means of
$|\pscal{\phi_-,\chi}|\leq \norm{\phi_-}_{L^2}\norm{\cP^0_-\chi}_{L^2}\leq\kappa_8\norm{\phi_-}_{L^2}\norm{\chi}_{L^2}/(c_nR),$
we conclude that there exists a constant $\epsilon>0$ depending only on $\Lambda_0$, $\nu$, $N$ and the chosen space $W$, such that for $R$ large enough,
$\pscal{\cD_i^n\phi,\phi}\leq (g_0(0)-\epsilon+O(c_n^{-1/2}))\norm{\phi}_{L^2}^2.$
This ends the proof of Lemma \ref{estim_vp_above}.
\qed \end{proof}

\begin{lemma}\label{bd_H1}
Each $\Phi^n=(\phi_i^n,...,\phi_N^n)$ is bounded in $H^1(\R^3,\C^4)^N$ as $n\to\ii$.
\end{lemma}
\begin{proof}
We adapt arguments from \cite{ES2}. Using the self-consistent equation \eqref{eq_orbitales} and estimates similar to those of the proof of Lemma \ref{estim_vp_above}, one can prove that there exists a constant $\ell>0$ (independent of $c_n$, but depending on $N$, $\nu$ and $\Lambda_0$) such that
$$\tr(|\cD^0|^2\gamma^n)\leq N\,g_0(0)^2+\ell\tr((-\Delta)\gamma^n)+\ell c_n^2\tr((-\Delta)\gamma^n)^{1/2}.$$
Using now \eqref{estim_D0_c}, we infer that 
$$\tr(|\cD^0|^2\gamma^n)\geq N\,g_0(0)^2+\kappa c_n^2\tr((-\Delta)\gamma^n),$$
which shows that
$\sum_{i=1}^N\norm{\nabla\phi_i^n}_{L^2}^2=\tr((-\Delta)\gamma^n)$
is bounded.
\qed \end{proof}

The sequences $(\phi_i^n)$ being bounded in $H^1(\R^3,\C^4)$ for all $i=1,...,N$, we can now rewrite the self-consistent equation \eqref{eq_orbitales} as
\begin{equation}
\left(c \tilde g_1'(0)\alp\cdot p+ \tilde g_0(0)c^2\beta+\left(\rho_{\gamma^n}-\nu\right)\ast\frac1{|\cdot|}-\frac{\gamma^n(x,y)}{|x-y|}\right)\phi_i^n=\mu_i^n\phi_i^n+\epsilon_i^n
\label{eq_simplifiee}
\end{equation}
where $\lim_{n\to\ii}\norm{\epsilon_i^n}_{H^{-1}(\R^3,\C^4)}=0$ and by \eqref{prop_g}. We now apply the method of \cite{ES2} to conclude that $(\phi_1^n,...,\phi_N^{n})$ converges towards $(\phi_1,...,\phi_N)$ with $\phi=\left(^{\psi}_0\right)$, $\psi\in H^1(\R^3,\C^2)^N$ which is a solution of the Hartree-Fock equations, and that
\begin{equation}
\lim_{n\to\ii}\{\E^\nu_{1,c_n,c_n\Lambda_0}(\gamma^{n})-N\,g_0(0)\}=\E_{\rm HF}^\nu(\gamma_\psi).
\label{limit_electrons}
\end{equation}
By the estimates of the proof of Lemma \ref{estim_vp_above} and \eqref{estim_vacuum}, we deduce that
$$\lim_{n\to\ii}\left(2D(\rho_{\gamma^{n}},\rho_{Q^n_{\rm vac}})-2\Re\iint_{\R^6}\frac{\gamma^{n}(x,y)\cdot\overline{Q^n_{\rm vac}(x,y)}}{|x-y|}dx\,dy-D(\nu,\rho_{Q^n_{\rm vac}})\right)=0,$$
i.e. that the interaction between the vacuum and the rest of the system vanishes.
Since 
$\E^0_{1,c_n,c_n\Lambda_0}(Q^n_{\rm vac})\geq0$
by the stability of the free vacuum, we finally obtain the lower bound \eqref{limit_inf}. This ends the proof of \eqref{limit_energy_nonrel}.

\bigskip

\noindent\textbf{Step 5:} \textit{Conclusion}. For $c$ large enough $|\cD^0|\geq g_0(0)$ hence by Lemma \ref{encadrement_min_prop}
$$(1-4/(c\pi))g_0(0)|q|-\frac12 D(\nu,\nu)\leq E^\nu(N)\leq g_0(0)|q|$$
This implies that, for $c$ large enough,
$E^\nu(N)> E^\nu(N-K)+E^0(K)$
for any $K<0$ or $K>N$. On the other hand, it is well-known \cite{LS,Lions-87} that
$$E^\nu_{\rm HF}(N)<\min\{E^\nu_{\rm HF}(N-K)+E^0_{\rm HF}(K),\ K=1,...,N\}.$$
This proves that $(H_1')$ holds for $c$ large enough, by \eqref{limit_energy_nonrel}. Thus there exists a minimizer $Q_c$ for $E^\nu(N)$. It then suffices to apply again the analysis of Steps 3-4 to show that $Q_c$ satisfies \eqref{eq_nonrel}, and obtain the stated convergence of the electronic orbitals towards a minimizer of the Hartree-Fock energy. This ends the proof of Theorem \ref{non-rel}.\qed

%%%%%%%%%%%%%%%%%%%%%%%%%%%%%%%%%%%%
%%%%%%%%%%%%%%%%%%%%%%%%%%%%%%%%%%%%
\appendix
\section{Proof of Lemma \ref{lemma_continuity}}
We write as usual $Q=Q^{++}+Q^{--}+Q^{+-}+Q^{-+}$, where by assumption $Q^{++},Q^{--}\in\gS_1(\gH_\Lambda)$ and $Q^{+-},Q^{-+}\in\gS_2(\gH_\Lambda)$.
First, using \eqref{def_rho} we see that $\norm{\rho_Q}_{L^2}\leq \kappa_\Lambda\norm{Q}_{\gS_2(\gH_\Lambda)}$
for some constant $\kappa_\Lambda=O(\Lambda^{3/2})$. Therefore, we now only estimate $D(\rho_Q,\rho_Q)$ in terms of $\norm{Q}_{1;\cP^0_-}$.
The diagonal terms are treated as follows
\begin{multline*}
\int_{B(0,\Lambda)}\frac{|\widehat{\rho_{Q^{++}}}(k)|^2}{|k|^2}dk\leq \norm{\widehat{\rho_{Q^{++}}}}^2_{L^\ii}\int_{B(0,\Lambda)}|k|^{-2}dk\\
\leq(2\pi) ^3(4\pi\Lambda)\norm{\rho_{Q^{++}}}_{L^1(\R^3)}^2\leq (2\pi) ^3(4\pi\Lambda)\norm{Q^{++}}_{\gS_1(\gH_\Lambda)}^2.
\end{multline*}
For the off-diagonal terms, we use ideas from \cite[p. 540--547]{HLS1}. Let $\zeta$ be a function in $\cC'\cap \gH_\Lambda$. We compute
\begin{align*}
 |\pscal{\rho_{Q^{+-}},\zeta}|=|\tr(Q^{+-}\zeta)|&=|\tr(Q^{+-}\cP^0_-\zeta\cP^0_+)|\\
&\leq \norm{Q^{+-}}_{\gS_2(\gH_\Lambda)}\norm{\cP^0_-\zeta\cP^0_+}_{\gS_2(\gH_\Lambda)}.
\end{align*}
Let us now fix some $\alpha_0<4/\pi$. Similarly to \cite[Lemma 12]{HLS1}, it can be proved that there exists a positive constant $\kappa$ independent of $\alpha$ (but depending on $\Lambda$ and $\alpha_0$) such that
$\tr_{\C^4}(\cP^0_-(p)\cP^0_+(q))\leq \kappa|p-q|^2.$
Therefore 
$$\norm{\cP^0_-\zeta\cP^0_+}_{\gS_2(\gH_\Lambda)}\leq \kappa'\norm{\nabla\zeta}_{L^2}^2= 4\pi\kappa'\norm{\zeta}_{\cC'}^2\ \text{and}\ \norm{\rho_{Q^{+-}}}_\cC\leq \kappa''\norm{Q^{+-}}_{\gS_2(\gH_\Lambda)}$$
for some constant $\kappa''$ depending only on $ \Lambda$ and $\alpha_0$.\qed

\section{On the Structure of the Variational Set}
In this section, we consider an infinite-dimensional Hilbert space $\gH$ and a reference orthogonal projector $\Pi$ on $\gH$ such that $\Pi$ and $1-\Pi$ are both of infinite rank. We introduce $\gH_+:=(1-\Pi)\gH$ and $\gH_-:=\Pi\gH$. First we prove a useful reduction for projectors belonging to $\Pi+\gS_1^{\Pi}(\gH)$ (i.e. to $\Pi+\gS_2(\gH)$ by \cite[Lemma 2]{HLS1}). This decomposition is valid in a more general setting (for any Fredholm pair of projections $(P,\Pi)$ \cite{ASS}) but for the sake of simplicity, we restrict ourselves to the Hilbert-Schmidt case needed in the article. Then, we deduce a general structure result for the variational set
\begin{equation}
\cQ:=\left\{Q\in\gS^{\Pi}_1(\gH)\ |\ Q^*=Q,\ -\Pi\leq Q\leq 1-\Pi\right\}.
\label{v_set}
\end{equation}
\begin{theorem}[Projections and BDF-states in the Fock space]\label{reduction}
Let $P$ be an orthogonal projector on $\gH$ such that $P-\Pi\in\gS_2(\gH)$. Denote by $(f_1,...,f_N)\in (\gH_+)^N$ an orthonormal basis of $E_{1}=\ker(P-\Pi-1)=\ker(\Pi)\cap\ker(1-P)$ and by $(g_1,...,g_M)\in (\gH_-)^M$ an orthonormal basis of $E_{-1}=\ker(P-\Pi+1)=\ker(1-\Pi)\cap\ker(P)$. Then there exist an orthonormal basis $(v_i)_{i\geq1}\subset \gH_+$ of $(E_{1})^\perp$ in $\gH_+$, an orthonormal basis $(u_i)_{i\geq1}\subset \gH_-$ of $(E_{-1})^\perp$ in $\gH_-$, and a sequence $(\lambda_i)_{i\geq 1}\in \ell_2(\R^+)$ 
such that
\begin{equation}
P=\sum_{n=1}^N|f_n\rangle\langle f_n|+\sum_{i=1}^\ii\frac{|u_i+\lambda_iv_i\rangle\langle u_i+\lambda_iv_i|}{1+\lambda_i^2},
\label{reduc_P}
\end{equation} 
\begin{equation}
1-P=\sum_{m=1}^M|g_m\rangle\langle g_m|+\sum_{i=1}^\ii\frac{|v_i-\lambda_iu_i\rangle\langle v_i-\lambda_iu_i|}{1+\lambda_i^2}.
\label{reduc_PP}
\end{equation} 
The Bogoliubov-Dirac-Fock state \cite{HLS1} associated with $P$ in the Fock space $\cal F$ built on the electron-positron decomposition $\gH=\gH_+\oplus\gH_-$ is then given by
\begin{eqnarray}
\Omega_P & = & k\prod_{n=1}^Na_0^*(f_n)\prod_{m=1}^Mb_0^*(g_m)\exp(Aa^*b^*)\;\Omega_0\label{formula_BDF_state}\\
 & = & k\prod_{n=1}^Na_0^*(f_n)\prod_{m=1}^Mb_0^*(g_m)\prod_{i\geq1}\big(1+\lambda_ia_0^*(v_i)b_0^*(u_i)\big)\;\Omega_0\label{formula_BDF_state2}
\end{eqnarray} 
where $A:=\sum_{i\geq1}\lambda_i|v_i\rangle\langle u_i|$ and $k=\prod_{i\geq 1}(1+\lambda_i^2)^{-1/2}$.
\end{theorem} 
 
Formula \eqref{formula_BDF_state} is classical and can be found in different forms in \cite{Thaller,KS,Ruis,SS} (see also \cite[Theorem 2.2]{BLS}).
Since $$\Pi=\sum_{m=1}^M|g_m\rangle\langle g_m|+\sum_{i\geq 1}|u_i\rangle\langle u_i|,$$
 we obtain
\begin{align}
P-\Pi & =  \sum_{n=1}^N|f_n\rangle\langle f_n|-\sum_{m=1}^M|g_m\rangle\langle g_m|+\sum_{i\geq 1}\frac{\lambda_i^2}{1+\lambda_i^2}\big(|v_i\rangle\langle v_i|-|u_i\rangle\langle u_i|\big)\label{expand_P_1}\\
 &  \qquad\qquad +\sum_{i\geq 1}\frac{\lambda_i}{1+\lambda_i^2}\big(|u_i\rangle\langle v_i|+|v_i\rangle\langle u_i|\big)\label{expand_P_3}.
\end{align} 
The terms in \eqref{expand_P_1} form the diagonal part of $P-\Pi$, which is trace-class \cite[Lemma 2]{HLS1}. The last term \eqref{expand_P_3} is the off-diagonal term which is only Hilbert-Schmidt \emph{a priori}. Note we obtain from this formula that
$\tr_{\Pi}(P-\Pi)=N-M$
is an integer \cite[Lemma 2]{HLS1}.
The formula of $P-\Pi$ can also be written as
$$P=\Pi+\sum_{n=1}^N|f_n\rangle\langle f_n|-\sum_{m=1}^M|g_m\rangle\langle g_m|+Q(A)$$
where $A:=\sum_{i\geq1}\lambda_i|v_i\rangle\langle u_i|$ and
$$Q(A)=\frac{A^*A}{1+A^*A}-\frac{AA^*}{1+AA^*}+A\frac{1}{1+A^*A}+\frac{1}{1+A^*A}A^*.$$
Therefore
\begin{multline}
\left\{P\ |\ P=P^*=P^2,\ P-\Pi\in\gS_2(\gH),\ \|P-\Pi\|<1\right\}\\
= \left\{ \Pi+Q(A),\ A\in\gS_2(\gH_+,\gH_-)\right\}.
\end{multline}

\begin{proof}[Proof of Theorem \ref{reduction}.] We only sketch the proof which is an easy adaptation of ideas in \cite{ASS,KS,Thaller,Kato1,Kato2,Ruis}. 
Let $U$ be a unitary transformation such that $P=U\Pi U^{-1}$. We introduce  
$U_{++}=(1-\Pi)U(1-\Pi)$, $U_{+-}=(1-\Pi)U\Pi$, $U_{-+}=\Pi U(1-\Pi)$ and $U_{--}=\Pi U\Pi$. It can be verified that $U_{+-}$ and $U_{-+}$ are Hilbert-Schmidt operators, and that $E_{1}=\ker U_{++}^*$ and $E_{-1}=\ker U_{--}^*$.
The operator $U_{--}:\ker(U_{--})^\perp\to {\rm Ran}(U_{--})=\ker(U_{--}^*)^\perp=E_{-1}^\perp$ possesses an inverse $U_{--}^{-1}$ well-defined and bounded on $E_{-1}^\perp$. Following \cite[Equation $(10.84)$]{Thaller}, we introduce the Hilbert-Schmidt operator $A:=U_{+-}U_{--}^{-1}:E_{-1}^\perp\to E_{1}^\perp$. It can be proved that
${\rm Ran}(P)=E_{1}\oplus^\perp (1+A)(E_{-1}^\perp)$
which means that $(E_1)^\perp$ is the graph of $A$. Writing
$A=\sum_{i\geq1}\lambda_i|v_i\rangle\langle u_i|$
where $(\lambda_i)_{i\geq1}\in \ell_2(\R^+)$, one obtains that
$(E_1)^\perp=(1+A)(E_{-1}^\perp)={\rm span}\{u_i+\lambda_i v_i,\ i\geq1\},$
and therefore
$$P=\sum_{n=1}^N|f_n\rangle\langle f_n|+\sum_{i=1}^\ii\frac{|u_i+\lambda_iv_i\rangle\langle u_i+\lambda_iv_i|}{1+\lambda_i^2}.$$
The same argument applies to \eqref{reduc_PP}.
The proof that, in the Fock space based on the decomposition $\gH=\gH_+\oplus\gH_-$,  the dressed vacuum $\Omega_P$  is given by formula \eqref{formula_BDF_state} is let to the reader.
Recall that $\Omega_P$ is characterized by the normalization constraint $\norm{\Omega_P}_{\cal F}=1$ and the relations $a_P(f)\Omega_P=b_P(f)\Omega_P=0$ for all $f\in\gH$, where $a_P(f)=a_0((1-P)f)+b_0^*((1-P)f)$ and $b_P(f)=a_0^*(Pf)+b_0(Pf)$.
\qed \end{proof}

We can now clarify the structure of the variational set $\cQ$ defined in \eqref{v_set}.
\begin{theorem}[Structure of the Variational Set]\label{thm_form_v_set} The set $\cQ$ coincides with the set containing all the operators of the form
\begin{equation}
Q=U_D(\Pi+\gamma)U_{-D}-\Pi
\label{form_v_set}
\end{equation}
where
\begin{enumerate}
\item $D\in\gS_2(\gH)$ is such that $\ker D\supseteq\ker\Pi$ and $\ker D^*\supseteq\ker(1-\Pi)$;
\item $U_D=\exp(D-D^*)$;
\item $\gamma\in\gS_1(\gH)$ is a self-adjoint and trace-class operator such that $[\gamma,\Pi]=0$ and, denoting $\gamma^{--}=\Pi\gamma\Pi$ and $\gamma^{++}=(1-\Pi)\gamma(1-\Pi)$, then $-\Pi\leq\gamma^{--}\leq0$ and $0\leq\gamma^{++}\leq 1-\Pi$. 
\end{enumerate}
%Moreover, if $\tr_{\Pi}(Q)\geq0$ then $-\Pi<\gamma^{--}$ and if $\tr_{\Pi}(Q)\leq0$ then $\gamma^{++}<1-\Pi$.
\end{theorem}

\begin{proof}
Notice first that any $Q$ of the form \eqref{form_v_set} belongs to $\cQ$. Indeed $U_D\gamma U_{-D}\in\gS_1(\gH)$ and $U_D\Pi U_{-D}-\Pi$ is a difference of two orthogonal projectors which is in $\gS_2(\gH)$ since $D\in\gS_2(\gH)$ and therefore belongs to $\gS_1^\Pi(\gH)$ by \cite[Lemma 2]{HLS1}. The constraint $-\Pi\leq Q\leq 1-\Pi$ is obviously satisfied.
We now prove that any $Q\in\cQ$ can be written as in \eqref{form_v_set}.

\begin{lemma}\label{lemma_interm_v_set}
For any $Q\in\cQ$, there exists an orthogonal projector $P$ and a trace-class operator $\gamma'$ such that $[P,\gamma']=0$ and
\begin{equation}
Q=P-\Pi+\gamma'.
\label{decomp_interm_v_set}
\end{equation}
Moreover, $P$ and $\gamma'$ can be chosen such that $\tr_\Pi(P-\Pi)=0$.
\end{lemma}

\begin{proof}[Proof of Lemma \ref{lemma_interm_v_set}] Let be $Q\in\cQ$. Since $Q$ is compact, the essential spectrum of $Q+\Pi$ is $\{0,1\}$. We write $Q+\Pi=\chi_{(1/2;1]}(Q+\Pi)+\gamma'$ and show that $\tr|\gamma'|<\ii$, which will prove \eqref{decomp_interm_v_set}.
We can find an orthonormal basis $(\phi_n)_{n\geq1}\cup (\psi_n)_{n\geq1}$ of $\gH$ such that
$$Q+\Pi=\sum_{n\geq 1}r_n|\phi_n\rangle\langle\phi_n| + \sum_{n\geq 1}(1-s_n)|\psi_n\rangle\langle\psi_n|$$
where $r_n\in[0;1/2]$, $s_n\in[0;1/2)$ and $\lim_{n\to\ii}r_n=\lim_{n\to\ii}s_n=0$. Computing
\begin{multline*}
\tr(Q^{++}-Q^{--}) = \sum_{n\geq1}r_n\norm{(1-\Pi)\phi_n}^2 + \sum_{n\geq1}(1-r_n)\norm{\Pi\phi_n}^2\\
+ \sum_{n\geq1}(1-s_n)\norm{(1-\Pi)\psi_n}^2 + \sum_{n\geq1}s_n\norm{\Pi\psi_n}^2
\end{multline*}
which is finite for $Q\in\gS_1^{\Pi}(\gH)$, we get
$$\sum_{n\geq1}\norm{\Pi\phi_n}^2<\ii,\quad \sum_{n\geq1}\norm{(1-\Pi)\psi_n}^2<\ii\text{ and } \sum_{n\geq1}(r_n+s_n)<\ii.$$
Then $\gamma'=\sum_{n\geq 1}r_n|\phi_n\rangle\langle\phi_n|- \sum_{n\geq 1}s_n|\psi_n\rangle\langle\psi_n|$ belongs to the trace class $\gS_1(\gH)$.
The above decomposition can be easily modified in order to ensure that $\tr_\Pi(P-\Pi)=0$. It suffices to add eigenstates of $Q+\Pi$ in $[0,1/2]$ to $\chi_{(1/2;1]}(Q+\Pi)$ if $\tr_\Pi(P-\Pi)<0$, or to remove eigenstates of $Q+\Pi$ in $(1/2,1]$ to $\chi_{(1/2;1]}(Q+\Pi)$ if $\tr_\Pi(P-\Pi)>0$.
\qed \end{proof}

Let us now write $Q=P-\Pi+\gamma'$ with $\gamma'\in\gS_1(\gH)$, $[\gamma',P]=0$ and $\tr_\Pi(P-\Pi)=0$. We apply Theorem \ref{reduction} to $P$ and find an operator $A=\sum_{i\geq1}\lambda_i|v_i\rangle\langle u_i|\in\gS_2(\gH)$ and two orthonormal systems $(f_1,...,f_N)\in(\gH_+)^N$ and $(g_1,...,g_N)\in(\gH_-)^N$ such that \eqref{reduc_P} and \eqref{reduc_PP} hold. Let us then introduce $e_{-k}=g_k$ and $e_k=f_k$ for $k=1,...,N$, $e_{-N-j}=u_j$ and $e_{N+j}=v_j$ for $j\geq 1$. We also define the Bogoliubov angles \cite{CIL} as
$\theta_k=\pi/2$ for $k=1,...,N$ and $\theta_{N+j}=\text{arccos}(1+\lambda_j^2)^{-1/2}$ for $j\geq1$. Then
$P=U_D\Pi U_{-D}$ with $D=\sum_{k\geq1}\theta_k|e_{-k}\rangle\langle e_k|$. Thus, introducing $\gamma=U_{-D}\gamma'U_D$, we obtain the result $Q=U_D(\Pi+\gamma)U_{-D}-\Pi$.
\qed \end{proof}

%
% Non-BibTeX users please use

%
\end{document}